\documentclass[a4paper,11pt]{article}
\pdfoutput=1 

\usepackage{jheppub} 
\usepackage{amsfonts,amssymb}
\usepackage{graphics,psboxit,amsmath} 
\usepackage{subfigure}
\usepackage{graphicx}
\usepackage{verbatim}
\usepackage{young}
\usepackage{array,multirow,hhline}
\usepackage{rotating}
\usepackage{nicefrac}
\usepackage{color}
\usepackage{tocbibind}
\usepackage[normalem]{ulem}
\usepackage{datetime}

\definecolor{markcolor2}{rgb}{1,0,0}

\def\be{\begin{equation}}
\def\ee{\end{equation}}
\def\ba{\begin{eqnarray}}
\def\ea{\end{eqnarray}}

\def\del{\partial}
\def\crbig{\\\noalign{\vspace{1.1mm}}}

\def\k{\kappa}
\def\r{\rho}
\def\a{\alpha}

\def\b{\beta}

\def\g{\gamma}

\def\d{\delta}
\def\D{\Delta}

\def\th{\vartheta}

\def\Om{\Omega}
\def\l{\lambda}

\def\cN{{\cal N}}

\def\no{\noindent}

\def\qq{\qquad}

\def\IR{\relax{\rm I\kern-.18em R}}

\def\inv{^{\raise.0ex\hbox{${\scriptscriptstyle -}$}\kern-.05em 1}}

\def \ha {{\frac{1}{2}}}

\def \ov {\over}

 \def\Xint#1{\mathchoice
 {\XXint\displaystyle\textstyle{#1}}%
 {\XXint\textstyle\scriptstyle{#1}}%
 {\XXint\scriptstyle\scriptscriptstyle{#1}}%
 {\XXint\scriptscriptstyle\scriptscriptstyle{#1}}%
 \!\int}
 \def\XXint#1#2#3{{\setbox0=\hbox{$#1{#2#3}{\int}$}
 \vcenter{\hbox{$#2#3$}}\kern-.5\wd0}}
 
 \def\dashint{\Xint-}

\usepackage{amsfonts}
\usepackage{amsmath}
\usepackage{amssymb}

\title{\boldmath
Gravity duals of ${\cal N}=2$ superconformal field theories with no electrostatic description}

\author[1]{P. Marios Petropoulos,}
\author[2,3]{Konstadinos Sfetsos,}
\author[4]{Konstadinos Siampos}

\affiliation[1]{Centre de Physique Th\'eorique,
Ecole Polytechnique, CNRS UMR 7644\\
91128 Palaiseau Cedex, France}
\affiliation[2]{Department of Mathematics, University of Surrey,
Guildford GU2 7XH, UK}
\affiliation[3]{Department of Engineering Sciences, University of Patras,
26110 Patras, Greece}
\affiliation[4]{M\'ecanique et Gravitation, Universit\'e de Mons,
7000 Mons, Belgique}

\emailAdd{marios.petropoulos@cpht.polytechnique.fr}
\emailAdd{k.sfetsos@surrey.ac.uk}
\emailAdd{konstantinos.siampos@umons.ac.be.}
\preprint{CPHT-RR052.0613\\
\vskip -.5 cm
\hfill DMUS-MP-13/17}

\abstract{We construct the first eleven-dimensional supergravity solutions, which are regular, have no smearing and possess
only $SO(2,4)\times SO(3)\times U(1)_R$ isometry.
They are dual to four-dimensional field theories with ${\cal N} = 2$ superconformal symmetry.
We utilise the Toda frame of self-dual four-dimensional Euclidean metrics with $SU(2)$ rotational symmetry.
They are obtained by transforming the Atiyah--Hitchin instanton under $SL(2,\mathbb{R})$ and are expressed in
terms of theta functions.
The absence of any extra $U(1)$ symmetry, even asymptotically,
renders inapplicable the electrostatic description of our solution.
}

\begin{document}
\maketitle
\flushbottom

\newcommand{\eqn}[1]{(\ref{#1})}

\clearpage

\def\baselinestretch{1.2}
\baselineskip 19 pt

\section{Introduction}

General solutions of eleven-dimensional supergravity as duals of ${\cal N}=2$
superconformal field theories were constructed in \cite{Lin:2004nb}.
The metric and form fields are given by
\begin{eqnarray}
\label{11dsolution}
&&\mathrm{d}s_{11}^2=\kappa_{11}^{\nicefrac{2}{3}}\mathrm{e}^{2\lambda}\left(4\mathrm{d}s^2_{\text{AdS}_5}+z^2\mathrm{e}^{-6\lambda}\mathrm{d}\Omega_2^2+\frac{4}{1-z\,
\partial_z\Psi}\,(\mathrm{d}\varphi+\omega)^2-\frac{\partial_z\Psi}{z}\,\gamma_{ij}\mathrm{d}x^i\mathrm{d}x^j\right)\,,
\nonumber\\
&& G_4=\mathrm{d}C_3=\kappa_{11}\,F_2\wedge\mathrm{d}\Omega_2\, ,
\end{eqnarray}
where the $\mathrm{AdS}_5$ and $S^2$ have unit radii and
\begin{eqnarray}
\label{11dsolutiondef}
&&\gamma_{ij}\mathrm{d}x^i\mathrm{d}x^j=\mathrm{d}z^2+\mathrm{e}^\Psi(\mathrm{d}x^2+\mathrm{d}y^2)\,\qquad \mathrm{e}^{-6\lambda}=-\frac{\partial_z\Psi}{z(1-z\,\partial_z\Psi)}\,,
\nonumber\\
&&
\omega = \omega_x \mathrm{d}x  + \omega_y \mathrm{d}y \ , \qquad \omega_x=\frac12\partial_y\Psi\,,\qquad \omega_y=-\frac12\partial_x\Psi\,,
\\
&&F_2=2 (\mathrm{d}\varphi+\omega)\wedge\mathrm{d}\big(z^3\mathrm{e}^{-6\lambda}\big)
+2 z\big(1-z^2\,\mathrm{e}^{-6\lambda}\big)\mathrm{d}\omega-\partial_z \mathrm{e}^\Psi\mathrm{d}x\wedge\mathrm{d}y\ .
\nonumber
\end{eqnarray}
Hence, these solutions boil down to determining the scalar function $\Psi(x,y,z)$.
It turns out that this obeys the continual Toda equation
\begin{equation}
\label{Toda2}
\left(\partial_x^2+\partial_y^2\right)\Psi+\partial_z^2\,\mathrm{e}^\Psi=0\ ,
\end{equation}
found in the context of continuum Lie algebras \cite{Saveliev}.
Demanding background regularity imposes appropriate boundary conditions given by
\begin{equation}
\label{bcToda}
z=0:\qquad \mathrm{e}^\Psi=\text{finite}\neq0\,,\qquad \partial_z\Psi=0\ .
\end{equation}
In fact the ratio $\displaystyle {\partial_z \Psi/ z}$ should stay finite as $z\to 0$.

Finding explicit solutions to this well posed mathematical problem has been so far possible only in two cases:
\begin{enumerate}
\item Whenever there in an extra $U(1)$ symmetry of the background allowing one to map \eqn{Toda2} with \eqn{bcToda}
to the Laplace equation \cite{Ward:1990qt} and the electrostatic analogue problem of a line charge distribution
over an infinite conducting plane \cite{Lin:2004nb}. The specifics of the line charge distribution has an one to one correspondence
with $\cN=2$ quiver gauge theories \cite{Gaiotto:2009gz}. Since this realisation a lot of works, notably
 \cite{Donos:2010va,ReidEdwards:2010qs,Aharony:2012tz}, have appeared in the literature, all of them assuming, to the
best of our knowledge, this extra $U(1)$ isometry which essentially makes the problem tractable by elementary means.
\item When it is possible to find separable solutions. Due to the non-linearity of the continual Toda equation this method is quite limited in finding
solutions in our context. Defining the complex coordinate $q=\frac12 (x+iy)$,
a separable solution is the product of a quadratic polynomial in $z$ and a function obeying Liouville's equation leading to
\begin{equation}
\label{Toda.Ex2}
\mathrm{e}^\Psi=c_3\frac{|\partial f|^2}{\left(1-c_3|f|^2\right)^2}\,\left(-z^2+c_1\,z+c_2\right)\,,
\end{equation}
where the $c_i$'s are real constants and $f=f(q)$ is a locally univalent meromorphic function.
\end{enumerate}
Evidently it is extremely important to find genuine solutions to \eqref{Toda2}, \eqn{bcToda} since that would allow to explore novel aspects of these theories.
It is the purpose of the present paper to find the first such solution in the literature that would fall outside the realm of electrostatic analogue
description or \eqn{Toda.Ex2}.

The idea of our construction is the following: First we recall that \eqn{Toda2} has appeared before in
the context of four-dimensional self-dual Euclidean metrics with a rotational Killing vector.
A particular solution corresponds to the Atiyah--Hitchin instanton metric \cite{ahgm,ah}.
This was originally formulated by utilising the $SU(2)$ invariance of the metric and the Darboux--Halphen system
\cite{halph} of differential equations arising from self-duality. What is directly useful for our purposes
is not this formulation but rather the rewriting  of the Atiyah--Hitchin metric in terms of the continual Toda potential which was found by an appropriately coordinate transformation in \cite{Olivier:1991pa}.
This provides a solution to \eqn{Toda2}, but the corresponding function $\Psi$ does not satisfy the boundary conditions \eqn{bcToda}.
We overcame this final obstacle by first using the extension of the construction of \cite{Olivier:1991pa} to arbitrary solutions of the Darboux--Halphen system
given in \cite{Finley:2010hs}.
Then we transform the Atiyah--Hitchin instanton under the $SL(2,\mathbb{R})$ covariance of the Darboux--Halphen system and by appropriately choosing
parameters of the transformation we obtain the required solution.

The organisation of this paper is as follows: In Sec. \ref{gravins} we make the connection of solutions of the continual Toda equation appropriate
for our eleven-dimensional background \eqn{11dsolution} to solutions of the Darboux--Halphen system that appears in the construction of
four-dimensional instantons with $SU(2)$ symmetry. In Sec. \ref{conTod} we explain how the boundary conditions for the Toda
potential in \eqn{bcToda} can be actually implemented in our case. In Sec. \ref{DH.solution} we actually construct our solutions in terms of theta functions
and check that indeed satisfy the appropriate boundary conditions. In Sec. \ref{conc} we present our conclusions.
We have also written App. \ref{Toda.AH.Appen} with all details of the construction of the Toda frame for the Atiyah--Hitchin metric.
In App. \ref{modfor} we have collected all necessary properties of modular forms and elliptic integrals for our construction.
Finally, in App. \ref{electro.Toda} we have reminded solutions which have an additional $U(1)$ isometry and have an electrostatic description.

\boldmath
\section{Gravitational instantons}
\unboldmath \label{gravins}
The purpose of this section is to provide the necessary review materials for our later construction.

\subsection{Hyper-K\"ahler geometries}

Let us consider a hyperk\"ahler manifold, which is a solution of Einstein vacuum equations with at least one isometry
and associated Killing vector $\xi=\xi^\mu\partial_\mu$.
If
\begin{equation}
\label{selfKilling}
\nabla_\kappa\xi_\lambda=\pm\frac12\sqrt{g}\,\varepsilon_{\kappa\lambda}{}^{\mu\nu}\,\nabla_\mu\xi_\nu\,,
\end{equation}
then $\xi$ is called translational Killing vector, otherwise rotational.
To be more concrete, we look at metrics of the form \cite{Gibbons:1979xm}
\begin{equation}
\label{GH}
\mathrm{d}\ell^2=V\,\left(\mathrm{d}\varphi+\omega_i\,\mathrm{d}x^i\right)^2+V^{-1}\,\mathrm{d}s^2\,,\qquad
\mathrm{d}s^2=\gamma_{ij}\,\mathrm{d}x^i\,\mathrm{d}x^j\,,\qquad i=1,2,3\,,
\end{equation}
where the fibering is defined along the action of the Killing vector $\xi=\partial_\varphi$, associated with this isometry.
If $\partial_\varphi$ is a translational Killing vector then we can always choose a coordinate system such that
\begin{equation}
\mathrm{d} V^{-1}=\pm\star_\gamma\mathrm{d}\omega\,,\qquad \gamma_{ij}=\delta_{ij}\,.
\end{equation}
For this metric the self-duality (or anti-self duality) condition can be simply written as
\begin{equation}
\label{self-trans}
\partial_i\partial^iV^{-1}=0\,,
\end{equation}
thus $V^{-1}$ is the ``electrostatic potential'' of the translational Killing vector. The localised solutions of \eqref{self-trans} read
\begin{equation}
\label{Trans.Laplace}
V^{-1}=\varepsilon+\sum_{i=1}^n\frac{m_i}{\vert \vec x-\vec x_{0i}\vert}\ ,
\end{equation}
where (multi-)Eguchi--Hanson and (multi-)Taub--Nut correspond to $\varepsilon=0,1$ respectively.
If $\partial_\varphi$ is a rotational Killing vector then we can always choose a coordinate system such that
\begin{equation}
\label{Boyer}
\mathrm{d}s^2=\mathrm{d}z^2+\mathrm{e}^\Psi(\mathrm{d}x^2+\mathrm{d}y^2)\,,\qquad
V^{-1}=\frac{1}{2}\partial_z\Psi\,,\qquad
\omega_x=\frac12\partial_y\Psi\,,\qquad \omega_y=-\frac12\partial_x\Psi\,,
\end{equation}
which is the Toda frame \cite{Boyer}.
For this metric the self-duality (or anti-self duality) condition is the three-dimensional continual Toda \eqref{bcToda} which here we write using complex coordinates $2q=x+iy$ and c.c.
\begin{equation}
\label{Toda.Equation}
\partial\bar\partial\Psi+\partial_z^2\,\mathrm{e}^\Psi=0\ ,
\end{equation}
The conformal transformations are symmetries of the Toda equation; i.e. $q\mapsto F(q)$ and
$\Psi\mapsto\Psi-\ln|\partial F|^2$ leaves invariant \eqn{Toda.Equation}.
Furthermore, the Kretschmann scalar of \eqref{GH}, \eqref{Boyer} has leading singular behaviour
\begin{equation}
\label{Riem.square}
R_{\kappa\lambda\mu\nu}\,R^{\kappa\lambda\mu\nu}\sim V^6\sim \left(\partial_z\Psi\right)^{-6}\,.
\end{equation}
Hence, regular solutions of four-dimensional Euclidean self-dual metrics require that $\partial_z\Psi\neq 0$.
This is not in conflict with the boundary condition \eqn{bcToda} since the latter refers to eleven-dimensional metrics.
However, this implies that we cannot simply take over solutions to Toda equations
appropriate for the four-dimensional metrics \eqn{GH} and use them in \eqn{11dsolution} since the resulting background
will be singular.

\subsection{Review of Bianchi IX folliations and Toda frame}

The way to find solutions to \eqn{Toda.Equation} is to make contact with another class of four-dimensional metrics of Euclidean Einstein gravity,
namely metrics of the Bianchi IX type
\begin{equation}
\label{metric.fol}
\mathrm{d}\ell^2=\frac{1}{4}\,(abc)^2\mathrm{d}t^2+a^2\sigma_1^2+b^2\sigma_2^2+c^2\sigma_3^2\, ,
\end{equation}
where $a,b,c$ are functions of $t$ and  $\sigma^i$ are the left-invariant Maurer--Cartan forms of $SU(2)$.
We choose our normalisation such that
\be
\mathrm{d}\sigma_i =\ha \varepsilon_{i jk}\sigma_j\wedge \sigma_k\ .
\ee
We will use the explicit parametrisation
\begin{equation}
\begin{array}{rcl}
&&\sigma_1+i\sigma_2=-\mathrm{e}^{i\,\psi}\left(i\,\mathrm{d}\vartheta+\sin\vartheta\,\mathrm{d}\varphi\right)\,,\quad
\sigma_3=\mathrm{d}\psi+\cos\vartheta\,\mathrm{d}\varphi\,,\crbig
&&{\rm Vol}(S^3)= \sigma_1\wedge\sigma_2\wedge\sigma_3 =
\sin\vartheta\,\mathrm{d}\vartheta\wedge\mathrm{d}\varphi\wedge\mathrm{d}\psi\, ,
\end{array}
\end{equation}
where for later use we have also written the volume form of the three-sphere.
The range of variables is
\be
\psi\in[-2\pi,2\pi]\ , \qq \vartheta\in[0,\pi]\ ,\qq \varphi\in[0,2\pi] \ .
\ee

The Bianchi IX foliation \eqref{metric.fol} is invariant under the left-action of the $SU(2)$ algebra,
generated by the Killing (right-invariant) fields
\ba
  \label{LKil}
&& \xi_1=  -\cos\varphi \cot\vartheta\, \partial_\varphi-\sin \varphi \,\partial_\vartheta+\frac{\cos \varphi}{\sin \vartheta} \, \partial_\psi\ ,
\nonumber\\
&& \xi_2=  \sin\varphi \cot\vartheta\, \partial_\varphi-\cos \varphi\, \partial_\vartheta-\frac{\sin \varphi}{\sin \vartheta} \, \partial_\psi \ ,
 \\
&& \xi_3= \partial_\varphi \ .
\nonumber
\ea
This exhausts the isometry when $a\neq b\neq c$. When two metric coefficients are equal, an extra $U(1)\subset SU(2)$ (right-action) survives.
For $a=b$ e.g. this $U(1)$ is generated by $\partial_\psi$.

\no
The Killing fields \eqref{LKil} can be rotational or translational depending on the self-duality conditions. As discussed in the literature
in several
instances  \cite{Gibbons:1979xm, Taub-nut, Hawking:1976jb, Eguchi:1978gw, Eguchi:1979yx, Gibbons:1979zt,Gibbons:1979xn},\footnote{See also \cite{Eguchi:1980jx}
for a general review of the subject and for an exhaustive classification see also \cite{Bourliot:2009fr,Bourliot:2009ad}.} integrating once the self-duality conditions for the curvature leads
to the first order system of non-linear differential equations
\ba
\label{DHL}
\dot a=\frac{a}{4}\, (b^2+c^2 -a^2 -2\lambda b c) \ ,
\nonumber\\
\dot b=\frac{b}{4}\,(c^2+a^2 -b^2 -2\lambda c a)\ ,
\\
\dot c=\frac{c}{4}\,(a^2+b^2 -c^2 -2\lambda a b) \ ,
\nonumber
\ea
where the derivative is with respect to $t$.
As it turns out, the independent constant of integration, is the parameter $\l$ which can take values $0$ and $1$ (in the former case, the connection is self-dual). 
The value $\lambda=0$ corresponds to the Lagrange system and \eqref{LKil} are translational Killing vectors. The
value $\l=1$ corresponds to the Darboux--Halphen system and \eqref{LKil} are rotational Killing vectors.
Whenever a right $U(1)$ remains in the isometry group, i.e.
when two of the functions $a,b$ and $c$ are equal that this is of the opposite nature \cite{Gibbons:1987sp}.
For our purposes we shall focus on the case of the rotational Killing vector $\xi_3$ ($\lambda = 1$) for which we shall find its corresponding Toda frame.
After a lengthy computation which is presented in full detail in App. \ref{Toda.AH.Appen} we find that
\ba
\label{AH.Toda.fin}
 \mathrm{e}^{2\Psi}&=&4\Big[\left[a\,(b-c)\,\sin^2\vartheta+c\,(a-b)\,\left(\sin^2\psi-\cos^2\vartheta\cos^2\psi\right)\right]^2
\nonumber\\
&&+
c^2\,(a-b)^2\,\cos^2\vartheta\,\sin^22\psi\Big]\, .
\ea
This is still in terms of the original coordinates $\th,\psi$ and $t$. The appropriate coordinate transformation is given by
\begin{equation}
\begin{array}{rcl}
&&z=\frac{1}{2}\,\left(c(a+b)+\sin^2\vartheta\left(b(a-c)-c(a-b)\,\sin^2\psi\right)\right)\ ,
\crbig
&&q=\displaystyle{\frac{1}{4}\,P\cos\vartheta+\frac{i}{2\sqrt{2}}\left\{\sqrt{b(a-c)}\,E\left(i\,p,i\frac{\kappa'}{\kappa}\right)-
\frac{ab}{\sqrt{b(a-c)}}\,F\left(i\,p,i\frac{\kappa'}{\kappa}\right)\right\}}\ ,
\end{array}
\label{coordzq}
\end{equation}
where we have defined
\begin{equation}
\begin{array}{rcl}
&&\displaystyle{P^2=2(a(b-c)+c(a-b)\cosh^2p)\ ,\quad p=i\left(\psi+\nicefrac{\pi}{2}\right)+\ln\tan\frac{\vartheta}{2}}\ ,
\\
&& 
\displaystyle{ \k = \sqrt{b(c-a)\ov a(c-b)}\ ,\qq \k'=\sqrt{1-\k^2} = \sqrt{c(a-b)\ov a(c-b)}}\,
\end{array}
\end{equation}
and $F(t,k)$ and $E(t,k)$ are the incomplete elliptic integrals of the first and second kind, respectively which we write for
completeness
\begin{equation}
F(t,k)=\int_0^t\mathrm{d}y\frac{1}{\sqrt{1-k^2\sin^2y}}\ ,\qquad E(t,k)=\int_0^t\mathrm{d}y\sqrt{1-k^2\sin^2y}\ .
\end{equation}

The only task remaining is to find appropriate solutions of the Darboux--Halphen system for the $a,b$ and $c$ and use them to construct
the Toda potential $\Psi$ from \eqn{AH.Toda.fin}. The choice of these functions should be such that the boundary conditions \eqn{bcToda}
are satisfied. This is a non-trivial task which we will undertake in the next section.

\boldmath
\section{${\cal N}=2$ SCFTs and continual Toda} \label{conTod}
\unboldmath
In this section we will construct a family of solutions to the continual Toda equation which will respect the appropriate
boundary conditions \eqn{bcToda} for the eleven-dimensional solution \eqn{11dsolution}.
As we shall see this will be done by using the Atiyah--Hitchin instanton solution,
followed by an appropriate $SL(2,\mathbb{R})$ transformation which leaves the Darboux--Halphen system
invariant.

We begin by first expressing various terms in the eleven-dimensional supergravity background \eqn{11dsolution} with the
definitions \eqn{11dsolutiondef} in terms of the variables $\th,\psi$ and $t$ as we have already done for the Toda potential $\Psi$ in \eqn{AH.Toda.fin}
and the coordinates $z$ and $q$ in \eqn{coordzq}. The information contained until \eqn{gijVV} is enough to construct the eleven-dimensional background
solution \eqn{11dsolution} once the solutions for $a,b$ and $c$ for the Darboux--Halphen system \eqn{DHL} (with $\l=1$) we construct in Sec. \ref{DH.solution}  are available.

First we compute from \eqn{Boyer}
\be
V =\sin^2\vartheta\,(a^2\cos^2\psi+b^2\sin^2\psi)+c^2\,\cos^2\vartheta\
\ee
and introduce (or recalling it from \eqn{nut.pot})
\be
b_{\rm nut}=c\,(a+b-c)-\sin^2\vartheta\,\left((a-c)(a+c-b)-(a-b)(a+b-c)\,\sin^2\psi\right)\ ,
\ee
such that $z=\ha (V+b_{\rm nut})$. Then we have that various combinations appearing in the background \eqn{11dsolution} and
in \eqn{11dsolutiondef} are conveniently expressed as
\begin{equation}
\label{11dsolutionAH}
\begin{array}{rcl}
&&  \displaystyle{\frac{\partial_z\Psi}{z}=\frac{2}{V z}\,, \qquad 1-z\,\partial_z\Psi=-\frac{b_{\rm nut}}{V}\,,}
\crbig
&&\displaystyle{\mathrm{e}^{-6\lambda}=\frac{2}{b_{\rm nut} z}\,,\qquad z(1-z^2\,\mathrm{e}^{-6\lambda})=-\frac{V z}{b_{\rm{nut}}}}\ .
\end{array}
\end{equation}
In addition, we have that
\begin{equation}
\label{gijVV}
\begin{array}{rcl}
&& \displaystyle{\g_{ij} \mathrm{d}x^i \mathrm{d}x^j = \mathrm{d} z^ 2 + e_+ e_- \ ,\qq \partial_z\mathrm{e}^\Psi\mathrm{d}x\wedge\mathrm{d}y=\frac{i}{V}\,e_+\wedge e_-}\ ,
\crbig
&&
\displaystyle{\omega=V^{-1}\left((b^2-a^2)\,\sin\vartheta\,\sin\psi\,\cos\psi\,\mathrm{d}\vartheta+c^2\,\cos\vartheta\,\mathrm{d}\psi\right)}\ ,
\end{array}
\end{equation}
where the components of the metric $\g_{ij}$ in the $(t,\vartheta,\psi)$ coordinate system can be found in \eqn{metric.Geroch}
as well as the frame components $\mathrm{e}_\pm$ in \eqn{infgr3}.
Our solution for $\Psi$ should satisfy \eqn{bcToda} or, equivalently the boundary conditions
\begin{equation}
\label{Toda.bc}
z\to0\,,\qquad V\to\pm\,\infty\,,\qquad \mathrm{e}^{\Psi}=\textrm{finite}\neq0\ ,
\end{equation}
and the product $V z$ be kept finite and different than zero.

\subsection{Appropriate boundary conditions}

We will explore the boundary condition \eqn{Toda.bc}.
In order to present the general analysis it is convenient to adopt an index notation. In particular, we will use $ a_i=(a,b,c)$ and the directional
cosines
\be
n_i=(\cos\psi\sin\vartheta,\sin\psi\sin\vartheta,\cos\vartheta)\ ,\qq \sum_{i=1}^3 n_i^2 =1 \ .
\ee
Then the Darboux--Halphen system, i.e. \eqn{DHL} for $\l=1$, can be written compactly as
\begin{equation}
\label{DH.Omega2}
\dot a_k= {a_i\ov 4} \left( (a_i - a_j)^2 - a_k^2\right) \,,\qquad {\rm cyclic\, in}\, (i,j,k)\ .
\end{equation}
An alternative to the $a_k$'s basis is to use the $\Omega_k$'s defined as
\be
\Omega_k =a_i a_j \ ,\qq a_k^2 = {\Omega_i \Omega_j\ov \Omega_k}\ ,\qq  {\rm cyclic\, in}\, (i,j,k)\ ,
\label{omaa}
\ee
where we also wrote the inverse relation.
They satisfy
\begin{equation}
\label{DH.Omega}
\dot\Omega_k=\ha \left(\Omega_i\Omega_j-\Omega_k(\Omega_i+\Omega_j)\right)\,,\qquad {\rm cyclic\, in}\, (i,j,k)\ ,
\end{equation}
which is the original Darboux system \cite{Darboux}.
Then note that
\ba
&& \mathrm{e}^{2\Psi}=4\sum_{i,j=1}^3\Omega_i\Omega_j\left(n_i^2+n_j^2+n_i^2n_j^2-1-2(2n_in_j-1)\delta_{ij}\right)\ ,
\\
&&
z =\frac12\sum_{i=1}^3\Omega_i\left(1-n_i^2\right)\, ,\quad V=\sum_{i=1}^3a_i^2n_i^2=
\Omega_1\Omega_2\Omega_3\sum_{i=1}^3 \left(\frac{n_i}{\Omega_i}\right)^2
\, .
\ea
This rewriting is particularly convenient as it treats all three coordinate axes as equivalent.

\no
We return to the determination of the behaviour of the various functions necessary to satisfy the boundary conditions.
How this behaviour is actually achieved together with the full solution will be the subject of the next section.

The boundary conditions are easily described in terms of the $z$ coordinate. However, in terms of the
angles $\vartheta$ and $\psi$, equivalently the directional cosines, and the coordinate $t$ the description is more involved.
Since the solution for $\Psi$ is written explicitly in terms of these coordinates,
it is necessary to provide such a description as well.  Let's denote with a star the constant variables for which the boundary condition is
achieved, i.e.  $t_*$ and $n^*_{k}$.
Generically, if $n^*_k\neq 0$ (there is always such a directional cosine) the solution of the boundary conditions
requires that $a_k\to \infty$ for some $t\to t_*$. From that we see that there are values for the $\Om$'s which will be denoted by a star such that
as $t\to t_*$ they have the following specific limits
\be
\Omega_k^* =0 \ ,\quad \Omega_i^* , \Omega_j^* ={\rm finite}\neq 0 \qq \Longleftrightarrow \qq a_k\to \infty\ ,\quad a_i, a_j \to 0 \ .
\label{bc.general1}
\ee
These guarantees that $\mathrm{e}^\Psi = {\rm finite}\neq 0$ and that $V\to \infty$.
Next, vanishing of $z$ requires in addition that
\be
\label{bc.general}
(1-n^{*2}_i)\Omega_i^*+(1-n^{*2}_j)\Omega_j^*=0\ ,
\ee
which already reveals that we must have that $\Omega_i^* \Omega_j^* < 0$.
Let's also define for later convenience
\be
\alpha=2\frac{1-n^{*2}_i}{1+n^{*2}_k}\,,\qquad \beta=2\frac{1-n^{*2}_j}{1+n^{*2}_k}\ ,\qq \alpha+\beta=2\,,\qquad \alpha,\beta\in (0,1]\ .
\label{abdef}
\ee
Note that none of the $\a$ or $\b$ can become zero since in that case $n^*_k=0$ which would have violated our hypothesis that $n^*_k\neq 0$.
Then \eqn{bc.general} is equivalent to the condition
\be
\a \Omega_i^* + \b \Omega_j^* =0 \ .
\label{aombom}
\ee
Keeping $zV$ finite requires some extra care. We have that
\ba
 2z\,V&=& \Omega_i^*\Omega_j^*n^{*2}_k \lim_{t\to t_*} \left(\frac{(1-n_i^{*2})\Omega_i+(1-n_j^{*2})\Omega_j}{\Omega_k} \right) + \Omega_i^*\Omega_j^*n_k^{*2}(1-n_k^{*2})
\nonumber\\
&
=& \Omega_i^*\Omega_j^*n_k^2 \lim_{t\to t_*} \left(\frac{(1-n_i^{*2})\dot \Omega_i+(1-n_j^{*2})\dot \Omega_j}{\dot \Omega_k} \right)  + \Omega_i^*\Omega_j^*n_k^{*2}(1-n_k^{*2})
\nonumber \\
& 
=& - 2\Omega_i^*\Omega_j^* n_k^{*4} = {\rm finite}\neq 0\ ,
\ea
where we used that
\be
\lim_{t\to t_*}\dot \Omega_k = - \lim_{t\to t_*} \dot \Omega_i  = - \lim_{t\to t_*} \dot \Omega_j = \Omega_i^* \Omega_j^*\ .
\ee
Then this gives the smooth limit as $z\to0$
\begin{equation}
\frac{\partial_z\Psi}{z}=\frac{2}{z\,V}=-\frac{2}{\Omega_i^*\Omega_j^*\, n_k^{*4}}={\rm finite}\neq0\,.
\end{equation}
Thus the last term in the eleven-dimensional metric \eqref{11dsolution} remains finite \cite{Aharony:2012tz}.
In the following, we will use the $SL(2,\mathbb{R})$ transformations of the Atiyah--Hitchin instanton 
satisfying \eqref{bc.general1}, 
as a solution-generating pattern for eleven-dimensional supergravity.
The solutions, which will be regular, have no smearing and possess only $SO(2,4)\times SO(3)\times U(1)_R$ isometry.

\boldmath
\subsection{Punctures}
\unboldmath
The boundary conditions which were given in \eqref{bcToda} are satisfied in the regions with no punctures.
In the case of punctures, the $S_\varphi$ circle, associated to the $U(1)$, shrinks in a smooth manner if \cite{Lin:2004nb}
\begin{equation}
\mathrm{e}^\Psi\sim (z_c-z)\,,\qquad {\rm at}\ z=z_c\,.
\end{equation}
This condition is related with the existence of a non-trivial 4-cycle which can support a non-trivial 4-form flux.
More concretely, to have a non-trivial 4-cycle that support 4-flux, we look for solutions where $S_\varphi$ circle shrinks to zero size at a point $z=z_c$, such that the 4-cycle is given
by the product $S_\varphi\times D_z\times S^2$, where $D_z=\{z\in[0,z_c]\}.$ An important quantity which we can compute from the gravity dual turns out to the be the energy of an M2-brane wrapping this 2-cycle
\cite{Gaiotto:2009gz}
\begin{equation}
\label{Energy}
\Delta=-\frac{2\kappa}{(2\pi)^2\ell_p^3}\int_{S^2}\,\left(\partial_z\,\mathrm{e}^\Psi\right)\,\mathrm{d}x\,\mathrm{d}y=2(g-1)\,N+K_T\,,\quad K_T=\sum_i\,K_i\,,\quad\
\kappa=\frac{\pi\ell_p^3}{2}\,,
\end{equation}
where $g$ is the genus of the surface, $z=z_c$ corresponds to the zeros of the Toda potential
and $K_T$ is the number of punctures of the surface.
A rewriting of the integrant in terms of the quantities which were introduced in the App. \ref{Toda.AH.Appen} can be done with the use of \eqref{gijVV}.

We next apply this computation to the Toda potentials we have considered so far.

\no
\underline{Backgrounds with no electrostatic description:} In this case there is no extra $U(1)$ symmetry.
The 2-cycle is defined by the zeros of $\mathrm{e}^\Psi$, which  were given by the four different cases in \eqref{AH.Toda.Zeros}.
Out of them the last one with $a=b=c$ is not consistent with our boundary conditions and is dismissed. For all of the other three we obtain that
$\D=0$ and therefore a two-torus, i.e. $g=1$, with no punctures.

\no
\underline{$\text{AdS}_7\times S^4$ metric:} In this case the Toda potential is given by \eqn{Toda.AdS} , i.e. $\mathrm{e}^\Psi=\coth^2\zeta\neq0$. Hence,
there are no punctures and $g=1$.

\no
\underline{Maldacena--N\'u\~nez metric \cite{Maldacena:2000mw}:}
Using the Toda potential of \eqref{Toda.MN} we find that the two circle
 is defined for $z=z_c=N$
and the energy reads
\begin{equation}
\Delta=2N\dashint_0^\infty dr {r\ov (1-r^2)^2}=-2N\ ,
\end{equation}
where we have defined the integral by its principal value.
Therefore this corresponds to a 2-sphere with no punctures.

\boldmath
\section{Bianchi IX instantons with strict $SU(2)$ isometry}
\label{DH.solution}
\unboldmath

In this section we explicitly construct solutions of the Darboux--Halphen system in terms of theta functions that indeed
satisfy the appropriate boundary conditions.

\subsection{The Darboux--Halphen system and its generic solutions}

Our starting point is the four-dimensional self-dual gravitational instantons of the type
\eqref{metric.fol} which, by utilising \eqn{omaa} we will write as
 \begin{equation}\label{metans}
\mathrm{d}\ell^2
= \Omega_1\Omega_2\Omega_3\, \mathrm{d}T^2
    +
    \frac{\Omega_2\Omega_3}{\Omega_1}\sigma_1^2+
    \frac{\Omega_3\Omega_1}{\Omega_2}\sigma_2^2+
    \frac{\Omega_1\Omega_2}{\Omega_3}\sigma_3^2\ ,
\end{equation}
where $T=t/2$. We will focus on the Darboux--Halphen system, \eqref{DHL} with $\lambda=1$,
 because in the case at hand, the isometry group is strictly reduced to the left $SU(2)$,
generated by the Killings \eqref{LKil}. These are rotational and will provide solutions of the continual Toda,
potentially eligible for higher-dimensional supergravity embeddings, dual to superconformal field theories. As we will show,
the required boundary conditions can indeed be reached and this is the central result of the present work.

The Omega's satisfy \eqn{DH.Omega} without the overall $1/2$ factor on the right hand side due to the
rescaling on $t$ as above. For convenience we repeat it here
\begin{equation}
\label{DH}
\Omega'_k=\Omega_i\Omega_j-\Omega_k(\Omega_i+\Omega_j)\ ,\qquad {\rm cyclic\, in}\, (i,j,k)\ ,
\end{equation}
where now the derivative is with respect to $T$.
Consider the same system in the complex plane: $\omega_i(z)$, $z\in \mathbb{C}$.
The general solutions of this system have the following properties \cite{halph, Takhtajan:1992qb}:
 \begin{itemize}
    \item The $\omega$'s are regular, univalued and holomorphic in a
    region with movable boundary i.e. a dense set of essential
    singularities). The location of this boundary accurately determines the
    solution.
    \item If $\omega_{i,0}(z)$ is a solution, then
\begin{equation}\label{slact}
 \omega_i (z) = \frac{1}{\left(c z + d  \right)^2} \omega_{i,0}
    \left( \frac{a z + b}{c z + d} \right) + \frac{c}{c z +
    d}\ , \quad  \left(\begin{matrix}
      a & b \\ c & d
    \end{matrix}
  \right) \in SL(2, \mathbb{C} )\ ,
\end{equation}
is another solution with singularity  boundary moved according to $\displaystyle z\to\frac{a z + b}{c z + d}$.
\end{itemize}

The fully anisotropic case is our main motivation here. In this case
no  algebraic first integrals exist and the general solution (see
\cite{halph, Takhtajan:1992qb, maciejewski95}) is expressed in terms
of quasimodular forms,  $\omega_i \in QM^1_2\left(\Gamma(2)\right)$,
where $\Gamma(2)$ is the level-2 congruence subgroup of $SL(2,\mathbb{Z})$ (the subset of elements of the form $\left(\begin{smallmatrix}
      a & b \\ c & d
    \end{smallmatrix}\right)
  =\left( \begin{smallmatrix}
      1 & 0 \\ 0 & 1
    \end{smallmatrix} \right) \ \mathrm{mod} \ 2 $).
Concretely
\begin{equation}\label{DH-sol}
 \omega_i (z)=
    - \frac{1}{2} \frac{\mathrm{d }}{\mathrm{d }z } \log \mathcal{E}_i (z)\ ,
\end{equation}
with $\mathcal{E}_i (z)$ triplet\footnote{\label{modtran}Notice their general transformations as generated by $ z\to -1/z$ and $ z+1$:
\begin{eqnarray}
 z\to- {1\ov z}&:&\quad \begin{pmatrix}
   \mathcal{E}_1&
   \mathcal{E}_2&
   \mathcal{E}_3
  \end{pmatrix} \to  z^2   \begin{pmatrix}
   \mathcal{E}_2 &
   \mathcal{E}_1 &
   -\mathcal{E}_3
  \end{pmatrix}\ ,
  \nonumber\\
   z\to z+1&:&\quad \begin{pmatrix}
   \mathcal{E}_1&
   \mathcal{E}_2&
   \mathcal{E}_3
  \end{pmatrix} \to  - \begin{pmatrix}
   \mathcal{E}_3 &
   \mathcal{E}_2 &
   \mathcal{E}_1
  \end{pmatrix}\ ,
\nonumber
 \end{eqnarray}
which can be verified by using \eqn{modtheta} and the explicit expressions
\eqn{halphsol} below.
}
of  \emph{holomorphic weight-2 modular forms} of  $\Gamma(2)$.
Again,  the $SL(2,\mathbb{C})$ action (\ref{slact}) generates new solutions $\left\{\omega_{i,0}\right\}\to \left\{\omega_i\right\}$
with a displaced set of singularities in $\mathbb{C}$, whereas the $SL(2,\mathbb{Z})\subset SL(2,\mathbb{C})$ acts as a permutation on $\omega$'s.

\no
Since our purpose is the description of gravitational backgrounds, we are interested in
real solutions of the real coordinate $T$ are obtained following the general pattern as
\begin{equation}
\Omega_\ell (T) = i  \omega_\ell(iT) = - \frac{1}{2}
\frac{\mathrm{d } }{\mathrm{d } T } \log \mathcal{E}_\ell (iT)\ .
\label{omtom}
\end{equation}
The reason for the choice $iT$ as the argument is that we get converging real functions.
According to (\ref{slact}), new real solutions are generated as\footnote{Note also the induced $SL(2,\mathbb{R})$ transformations 
on the Toda variables $z$ and ${\rm e}^{2\Psi}$:
\begin{eqnarray*}
&&z(T,\vartheta,\psi)=\frac{1}{\left(C T+ D  \right)^2}\,z_0 \left( \frac{A T + B}{C T + D},\vartheta,\psi \right)+\frac{C}{CT+D}\,,\nonumber\\
&&{\rm e}^{2\Psi(T,\vartheta,\psi)}=\frac{1}{\left(C T + D  \right)^4}\,{\rm e}^{2\Psi_0\left( \frac{A T + B}{C T + D},\vartheta,\psi \right)}\,.
\end{eqnarray*}}
\begin{equation}\label{slactR}
  \Omega_i (T) = \frac{1}{\left(C T + D  \right)^2} \Omega_{i,0}
    \left( \frac{A T + B}{C T + D} \right) + \frac{C}{C T +
    D}\ , \quad  \left(\begin{matrix}
      A & B \\ C & D
    \end{matrix}
  \right) \in SL(2, \mathbb{R})\ .
\end{equation}
Note that the entries of the $SL(2,\mathbb{C})$ and $SL(2,\mathbb{R})$ transformation matrices are related as
$a= A, d=D, c=-i C$ and $b=i B$, so that the determinant condition is preserved. In addition, note that
an $SL(2,\mathbb{R})$ transformation on the $\Om$'s never changes their relative position in the real line. This remains the
same as in the seed solution. Also note that the Darboux system \eqn{DH} does not allow a cross over behaviour for
its solutions. The reason is that, if for some finite $T$ two of the $\Om$'s, initially different, become equal,
then their derivatives at that point are equal as well. Hence these $\Omega$'s will subsequently remain equal.

A particular solution of the Darboux system \eqn{DH} is
the original Halphen solution \cite{halph}. In the present language,  it corresponds to
    \begin{equation}\label{halphsol}
\begin{cases}
     \mathcal{E}_{1,\mathrm{H}} = i\pi\theta_4^4 \\
     \mathcal{E}_{2,\mathrm{H}} = -i\pi \theta_2^4  \\
     \mathcal{E}_{3,\mathrm{H}} = -i\pi\theta_3^4
  \end{cases}
\quad \Longleftrightarrow
\quad \begin{cases}
    \omega_{1,\mathrm{H}} = \frac{\pi}{6i}\left(E_2-\theta_2^4 - \theta_3^4 \right) \\
    \omega_{2,\mathrm{H}} = \frac{\pi}{6i}\left(E_2+\theta_3^4 + \theta_4^4 \right)  \\
    \omega_{3,\mathrm{H}} = \frac{\pi}{6i}\left(E_2+\theta_2^4 - \theta_4^4\right)\ ,
  \end{cases}
    \end{equation}
where we have used standard notation for the Theta functions and the quasimodular form of weight two. Details are given in App. \ref{modfor}.
This is also  the solution found by Atiyah and Hitchin \cite{ahgm,ah} as the
Bianchi IX self-dual gravitational instanton relevant for describing the configuration
space of two slowly moving BPS $SU(2)$ Yang--Mills--Higgs
monopoles \cite{Manton:1981mp,Gibbons:1986df}.
This solution and the $SL(2,\mathbb{R})$-transformed ones, possess rotational Killing fields, given in \eqref{LKil},
as opposed to the strict-$SU(2)$-isometry family of solutions of the Lagrange system found earlier in \cite{Belinsky:1978ue}, where the same fields are translational.

\subsection{Boundary conditions and the continual Toda equation}
\label{bcmodular}
\subsubsection{General solution and boundary conditions}

\begin{figure}[!h]
\begin{center}
\includegraphics[height=5cm]{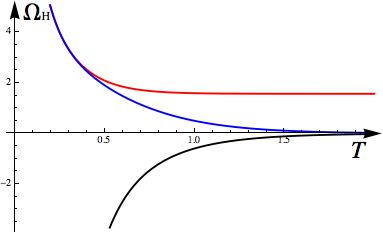}
\end{center}
\vskip -0.5 cm
\caption{Halphen original solution ($\Omega_{1,\mathrm{H}}<0<\Omega_{3,\mathrm{H}}<\Omega_{2,\mathrm{H}}$).} \label{halphorig}
\end{figure}

We now turn to the problem of finding solutions of the system \eqref{DH}, which satisfy the prescribed boundary condition that there exist
some $T=T_*$  so that the corresponding values for the $\Omega_i$'s, i.e. $\Om^*_i$'s satisfy \eqn{bc.general1} and \eqn{aombom}.
It should be noticed at this point that the metric ansatz \eqref{metans}
assumes the product of the three $\Omega$'s be positive. In the opposite instance, it is enough to change the overall sign of the ansatz to recover a positive-definite metric.
The domain of definition of the Euclidean time $T$ is set by the roots or the poles of the $\Omega$'s, and $T_¥$ which is a root of $\Omega_3$.
The existence of such a limiting value is consistently required from the higher-dimensional perspective,
even though from the point of view of the four-dimensional gravitational instanton, a simple root of any $\Omega_i$ is a genuine
curvature singularity (see \cite{Belinsky:1978ue}).
\no
The original Halphen solution \eqref{halphsol}, according to \eqn{omtom}, along the imaginary $z$-axis leads to real $\Omega$'s
\ba
 &&   \Omega_{1,\mathrm{H}} = \frac{\pi}{6}\left(E_2(iT)-\theta_2^4(iT) - \theta_3^4(iT) \right) \ ,
\nonumber \\
  &&  \Omega_{2,\mathrm{H}} = \frac{\pi}{6}\left(E_2(iT)+\theta_3^4(iT) + \theta_4^4(iT) \right)\ ,
\label{halphsolR}
\\
   && \Omega_{3,\mathrm{H}} = \frac{\pi}{6}\left(E_2(iT)+\theta_2^4 (iT)- \theta_4^4(iT)
    \right)\ .
\nonumber
\ea
These do not satisfy the boundary conditions \eqn{bc.general1} and \eqn{aombom}.
The reason is that none of the $\Omega$'s in that solution has any root (see Fig.\ref{halphorig}).
This precise property makes the Atiyah--Hitchin instanton regular everywhere in the range $0<T<\infty$
(connecting the Taubian infinity with the bolt).
\no
We will need the behaviours for small and large $T$ of the Halphen functions $\Omega_{i,\mathrm{H}}$ defined in \eqref{halphsolR}.
This is obtained from the well-known behaviours of the modular and quasimodular forms introduced above, along the imaginary axis $\tau=iT$ using the results of the
App. \ref{modfor}. We obtain that 
 \begin{equation}
 \label{sT}
T\to 0:  \quad\begin{cases}
 \Omega_{2\vert 3,\mathrm{H}} \approx\frac{1}{T}
 \\
 \Omega_{1,\mathrm{H}}\approx -\frac{\pi}{2T^2}
\end{cases}
\end{equation}
and that
 \begin{equation}
 \label{lT}
T\to \infty:
\quad\begin{cases}
 \Omega_{1\vert 3,\mathrm{H}}\approx \mp 4\pi\, {\rm e}^{-\pi T}
 \\
 \Omega_{2,\mathrm{H}}\approx \frac{\pi}{2} + 4\pi\,{\rm e}^{-2\pi T}
\end{cases}
\end{equation}


We will now show that a family of  solutions obtained from Halphen's  \eqref{halphsolR} using
the generating procedure \eqref{slactR} (with $\Omega_{i,0}=\Omega_{i,\mathrm{H}}$) can indeed satisfy \eqn{bc.general1} and \eqn{aombom}.
The first observation about \eqref{slactR} is that the $ \Omega$'s are defined only if
    \begin{equation}\label{conda}
    Z=\frac{AT+B}{CT+D} >0\ .
    \end{equation}
The reason for that is the argument of the various functions should have positive imaginary part for them to converge.
Let us introduce for later convenience
  \begin{equation}\label{param}
\begin{cases}
   T_0=-\frac{B}{A}\ \Leftrightarrow\ Z=0
    \\
    Z_0=\frac{B}{D}\ \Leftrightarrow\  T=0
  \end{cases}
\quad\text{and}\quad\
\begin{cases}
 T_\infty=   - \frac{D}{C}\ \Leftrightarrow\ Z\to \pm \infty\\
Z_\infty=    \frac{A}{C}\ \Leftrightarrow\ T\to \pm \infty\ .
  \end{cases}
    \end{equation}
There are two distinct generic behaviours for the $\Omega$'s,
depending on the sign\footnote{Note that $Z_\infty$ cannot vanish due to unimodularity of the trajectory. Furthermore $\displaystyle T_0=T_\infty+\frac{1}{AC}$ and consequently
\begin{itemize}
\item if $Z_\infty >0$, then $AC>0$ and $T_\infty<T_0\ \Leftrightarrow\ \frac{D}{C}>\frac{B}{A}$;
\item if $Z_\infty <0$, then $AC<0$ and $T_0<T_\infty\ \Leftrightarrow\ \frac{D}{C}<\frac{B}{A}$.
\end{itemize}
}
of $Z_\infty$:
\begin{description}
\item[$Z_\infty >0\ \Leftrightarrow\  \frac{D}{C}>\frac{B}{A}$:]  $\Omega$'s are defined on 2 disconnected sets $T<T_\infty<T_0<T$;
\item[$Z_\infty <0\ \Leftrightarrow\ \frac{D}{C}<\frac{B}{A}$:] $\Omega$'s are defined on a single set $T_0<T<T_\infty$ ($0<Z<\infty$).
\end{description}

\boldmath
\subsubsection{The regime $Z_\infty >0$}
\unboldmath

We will first prove that in the regime $Z_\infty >0$, condition \eqn{aombom} fails.
Let us concentrate on the region $T_0<T$ (central symmetry allows to draw similar conclusions for the alternative set $T<T_\infty$).
Using the results \eqref{sT} and \eqref{lT} of App. \ref{modfor}, we can study the behaviour of $\Omega$s for large $T$ and around $T_0$.

\begin{figure}[!h]
\begin{center}
\includegraphics[height=5.cm]{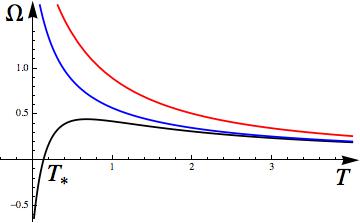}
\end{center}
\vskip -0.5 cm
\caption{Generic solution \eqref{slactR} for $Z_\infty>0$ in the range $T_0<T$
($\Omega_1<\Omega_3<\Omega_2$).}  \label{halphanis}
\end{figure}

\underline{Large-$T$ behaviour:}
This behaviour is dictated by the inhomogeneous term in \eqref{slactR}:
 \begin{equation}
\Omega_i = \frac{1}{T} + {\cal O}\left(\frac{1}{T^2}\right)\ ,
\end{equation}
with $0<\Omega_1<\Omega_3<\Omega_2$ (at large $T$, due to $\Omega_{1,\mathrm{H}}<\Omega_{3,\mathrm{H}}<\Omega_{2,\mathrm{H}}$ behaving as in \eqref{lT}).
\underline{Behaviour for $T_0 \lessapprox T$:} Around $T_0$, the argument of the $\Omega_{i,\mathrm{H}}$ vanishes. Using \eqref{sT} and \eqref{slactR} we obtain
  \begin{equation}
  \label{bT0}
     \Omega_1(T) \approx
    -\frac{\pi}{2A^2}\frac{1}{\left(T-T_0\right)^2}\ , \qquad \Omega_{2\vert 3}(T) \approx \frac{1}{T-T_0}.
\end{equation}
\no
It is not hard to interpolate between the large-$T$ and the $T_0 \lessapprox T$ regions, and obtain the global
picture for $\Omega_i(T)$. On the one hand, we observe that the $\Omega_{2\vert 3,\mathrm{H}}$ are positive and monotonically decreasing.
Using \eqn{slactR} we may compute that 
\be
{\mathrm d\Omega_i\ov \mathrm d T} = -2 {C^2\ov (T-T_\infty)^3} \Omega_{i,\mathrm H} + {1\ov C^2 (T-T_\infty)^2} \Omega'_{i,\mathrm H} - {1\ov (T-T_\infty)^2}\  .
\ee
Hence since $T>T_\infty$ a decreasing and positive $\Omega_{2\vert 3,\mathrm{H}}$ implies that  $\Omega_{2\vert 3}$ remains
decreasing and positive.
On the other hand, since $\Omega_1(T)$ is negative around $T_0 \lessapprox T$,
there is a $T_*$ such that $\Omega_1(T_*)=0$. We expect  $\Omega_1(T)$ to increase, become positive at $T_*$,
and decrease towards zero at large $T$. The generic behaviour is presented\footnote{From the four-dimensional perspective,
the corresponding gravitational instanton with metric \eqref{metans} has its asymptotic region around $T=T_0$, a nut charge at $T\to \infty$,
and a naked singularity at $T=T_*$ \cite{Petropoulos1}.} in Fig.\ref{halphanis}.
Since  $\Omega_{2\vert 3,\mathrm{H}}$ always stay non-negative there is no way we could satisfy the boundary condition
\eqn{aombom} (with $i=2$ and $j=3$).

 \boldmath
\subsubsection{The regime $Z_\infty <0$}
\unboldmath

Before entering the technical details, let us motivate the reason why the regime $Z_\infty <0$, or equivalently
\be
\frac{D}{C}<\frac{B}{A}\quad {\rm or}\quad T_0<T<T_\infty (0<Z<\infty)\ ,
\label{rangge4}
\ee
is appropriate for our purposes.
For this type of $SL(2,\mathbb{R})$ transformations, $T_0$ and $T_\infty$ are poles. The behaviour of the $\Omega$s around $T_0$ has already
been exhibited in \eqref{bT0}. Around $T_\infty$, we proceed in a similar manner.

\underline{Behaviour for $T \lessapprox T_\infty$:} Around $T_\infty$, the argument of the $\Omega_{i,\mathrm{H}}$ diverges. Using \eqref{lT} and \eqref{slactR} we find
  \begin{equation}
  \label{bTinf}
      \Omega_{1\vert 3}(T) \approx \frac{1}{T-T_\infty}\ , \quad  \Omega_2(T) \approx
    \frac{\pi}{2C^2}\frac{1}{\left(T-T_\infty\right)^2}\ .
\end{equation}

 \vskip -0.3 cm
\begin{figure}[!h]
\begin{center}
\includegraphics[height=5 cm]{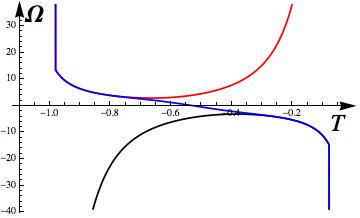}
\end{center}
\vskip -0.5 cm
\caption{Generic solution \eqref{slactR} for $Z_\infty<0$ in the range $T_0<T<T_\infty$
($\Omega_1<\Omega_3<\Omega_2$).}
\label{halphGM}
\end{figure}
\no
From the behaviours \eqref{bT0} and \eqref{bTinf}, together with the expression \eqref{slactR} and the known
behaviours of $\Omega_{i,\mathrm{H}}$, we can qualitatively describe the solutions at hand. The function $\Omega_3(T)$
interpolates between two simple poles. It is positive around $T_0$ and  negative around $T_\infty$, thus there is a root
at $T_*$, i.e.
\be
\Om^*_3 =\Omega_3(T_*) =  0\ ,
\label{eq1}
\ee
with  $\dot \Omega^*_3<0$. The function $\Omega_1(T)$ is finite and negative between its double pole
at $T_0$ and its simple pole at $T_\infty$, whereas $\Omega_2(T)$ is finite and positive between its simple pole at $T_0$ and
its double pole at $T_\infty$. The generic behaviour is depicted in Fig. \ref{halphGM} and this is all consistent with $\dot \Omega^*_3=\Omega_1^*\Omega_2^*<0$,
with $\Omega_1^*<0<\Omega_2^*$.
This analysis shows that for given $\alpha,\beta$ in \eqn{abdef}, we may indeed find $A,B,C,D$ within the range \eqn{rangge4} and
that \eqn{aombom} (with $i=1$ and $j=2$) is satisfied.  This is what we will precisely show now by computing $T_*$.

Using  \eqref{slactR} with $\Omega_{i,0}=\Omega_{i,\mathrm{H}}$, the explicit expressions of $\Omega_{i,\mathrm{H}}$ given in \eqref{halphsolR},
as well as condition \eqn{aombom} (for $i=1$ and $j=2$), the normalisation in \eqn{abdef} and the identity \eqn{theta4}, we obtain that
 \begin{equation} \label{ab}
\a = 2 {\th_4^4(i Z_*)\ov \th_3^4(i Z_*)}\ ,\qq  \b = 2 {\th_2^4(i Z_*)\ov \th_3^4(i Z_*)}\ ,
 \end{equation}
where
\be
Z_* ={A T_* + B \ov C T_*+D}  \ .
\label{cond}
\ee
This allows to determine $Z_*$ as an elliptic modulus, in terms of the complete elliptic integral
of the first kind $K(\kappa)$ introduced in \eqref{cel1}
 \begin{equation}
 \label{Zelmod}
Z_*  =\frac{K\left(\sqrt{\alpha/2}\right)}{K\left(\sqrt{\beta/2}\right)}
 \end{equation}
with the modulus and complementary modulus of the  elliptic integrals given by
 \begin{equation} \label{hgjjhgk}
\kappa^2 = \frac{\beta}{2}\ ,\qquad
\kappa'^2 =  \frac{\alpha}{2}
 \end{equation}
(see \eqref{elmod}--\eqref{elnorm}). Moreover, the asymptotic expression of $Z_*$ around
$\alpha=0,2$ reads
\begin{equation}
Z_*\simeq\frac{\pi}{\ln\frac{32}{\alpha}}\,,\qquad Z_*\simeq\frac{\ln\frac{32}{\beta}}{\pi}\,.
\end{equation}
Equation \eqref{eq1} allows finally to determine the critical value of $T$ as
  \begin{equation} \label{Tstar}
T_*=\frac{1}{\pi C^2}\left[
\alpha K\left(\sqrt{\beta/2}\right)
-2E\left(\sqrt{\beta/2}\right)
\right]K\left(\sqrt{\beta/2}\right)-\frac{D}{C}\ ,
 \end{equation}
where we have used the identities \eqref{id1} and  \eqref{id2} involving the complete elliptic integral of the second kind $E(\kappa)$.
Since $T_*$ and $Z_*$ are related via \eqref{cond}, $A,B,C,D$ are not arbitrary unimodular real numbers. They must satisfy
\be
{A\ov C} = - {\b K\left(\sqrt{\a/2}\right) - 2 E\left(\sqrt{\a/2}\right) \ov  \a K\left(\sqrt{\b/2}\right) - 2 E\left(\sqrt{\b/2}\right)}\ .
\label{eqS}
\ee
The approximate expressions near the extreme values $\alpha=0$ and $2$ read
\begin{equation}
\frac{A}{C}\simeq-\frac{\pi\alpha}{8}\ ,\qquad \frac{A}{C}\simeq\frac{8}{\pi(\alpha-2)}\ .
\end{equation}
At this stage the reader may be worried about inequality \eqref{rangge4} not being compatible with \eqref{eqS} combined with unimodularity
of the $SL(2,\mathbb{R})$  transformation
It is remarkable that owing to inequality \eqref{Cineq}, we learn from \eqref{eqS} that $A$ and $C$ have
always opposite signs and therefore, after multiplication by $AC$, \eqref{rangge4} reads
$
AD-BC>0
$. Thanks to unimodularity this is always valid.

The simplest possible choice is $\alpha=\beta=1$, because it is symmetric
and algebraic which is generically not the case.
Using \eqref{Zelmod} we find $Z_*=1$,
whereas  \eqref{eqS} leads to  $A+C=0$. We further obtain $T_*=\nicefrac{1}{2}\left(T_0+T_\infty\right)$, which is a centre of symmetry.

\boldmath
\section{Concluding remarks}\label{conc}
\unboldmath

The main result of this paper is the construction of the first in the literature solution of
eleven-dimensional supergravity as dual of field theories with ${\cal N} = 2$ superconformal symmetry which has
only $SO(2,4)\times SO(3)\times U(1)_R$ isometry.
Our construction was made possible by making contact with solutions of the continual Toda equation corresponding to the four-dimensional
Atiyah--Hitchin gravitational instanton and subsequent use of modular transformations in order to satisfy the appropriate boundary conditions.

Our solutions have no region in which a further $U(1)$ symmetry develops and therefore cannot be described in terms of rotationally invariant Laplace equation -- electrostatic description. 
A solution to the continual Toda equation can be reconstructed using solutions of the free part of the equation,
i.e. the two-dimensional Laplace equation (see \cite{Bakas:1996gf} and citations therein), which are called free fields.
However, this reconstruction is typically a perturbative infinite resummation based on various expectation values of 
these free fields. This is 
dictated by the underlying group-theoretical structure and
preserves the symmetries of the seed solution of the two-dimensional Laplace equation.
Hence, if we consider solutions of the Laplace equation in
the two-dimensional infinite plane with $U(1)$ symmetry, the solution which would be generated perturbatively  
would necessarily share the same $U(1)$
symmetry. An example of this is the reconstruction of the Toda potential for the Eguchi--Hanson metric in \cite{Bakas:1996gf} 
(see equation (4.11) and below in that
paper). The potential for the Maldacena--N\~un\'ez metric \eqn{Toda.MN} being just an analytic continuation of the Eguchi--Hanson one, this  indicates a deeper connection between the electrostatic methods developed in the
recent years with the free-field group-theoretical approach to solutions of the continual Toda equation used in the past.

A natural question is how to obtain a free-field realisation of solutions to the continual Toda equation with no extra $U(1)$ symmetry. One way to think of this is advocating non-perturbative
effects, which break the symmetry of a particular seed solution on the infinite plane. It seems, however, more systematic to seek for solutions of the
free equation on Riemann surfaces instead of the infinite plane. This might be related to an electrostatic analogue problem, not as simple as that of a
charged line intersecting the infinite conducting plane, but instead with a set up involving higher-genus surfaces and charge densities.

Finally, it will be of much interest to explore the underlying physical information of our solution with regard to field theory and keeping in mind the above comments.

\section*{Acknowlegments}
This work was initiated during the XLII\`eme Institut d'\'et\'e de l'\'Ecole Normale Sup\'erieure, in Paris in August 2012.
The authors have benefited from discussions with I. Bakas, J.P. Derendinger, D. Giatatanas, S. Katmadas, Ph. Spindel and D.C. Thompson.
The research of P.M. Petropoulos was  supported by the LABEX P2IO, the ANR contract 05-BLAN-NT09-573739, the ERC Advanced Grant 226371 and the ITN programme PITN-GA-2009-237920.
The research of K.\,Sfetsos is implemented under the \textsl{ARISTEIA} action of the \textsl{operational
programme education and lifelong learning} and is co-funded by the European Social
Fund (ESF) and National Resources.
The work of K. Siampos has been supported by  \textsl{Actions de recherche concert\'ees (ARC)} de la \textsl{Direction g\'en\'erale
de l'Enseignement non obligatoire et de la Recherche scientifique Direction de la Recherche scientifique Communaut\'e
fran\c{c}aise de Belgique}, and by IISN-Belgium (convention
4.4511.06). P.M. Petropoulos and K. Siampos would like to thank each others home institutions
and the Universities of Patras and Surrey for hospitality, where part of this work was developed.

\vfill

\appendix

\section{Toda frame of the Atiyah--Hitchin metric}
\label{Toda.AH.Appen}
\subsection{Construction of the Toda frame}
In this appendix we shall revisit the derivation of the Toda frame for the Atiyah--Hitchin metric as this was presented in \cite{Olivier:1991pa}
and generalised in \cite{Finley:2010hs} for the general solution of the Darboux--Halphen first order (DH) system.
We shall present the derivation done in \cite{Finley:2010hs}, adopted however in our conventions. The key point of this computation is the repeated
use of \eqref{DHL} for the Darboux--Halphen system, i.e. $\lambda=1$.

Rewriting of the metric \eqref{metric.fol} in the Geroch formulation with a Killing vector $\xi_3$
\begin{eqnarray}
\label{metric.Geroch}
&&\mathrm{d}\ell^2=V\,\left(\mathrm{d}\varphi+\omega_i\,\mathrm{d}x^i\right)^2+V^{-1}\,\mathrm{d}s^2\, ,\qquad
\mathrm{d}s^2=\gamma_{ij}\,\mathrm{d}x^i\,\mathrm{d}x^j\,,\qquad x^i=(t,\vartheta,\psi)\, ,
\nonumber\\
&&V=\sin^2\vartheta\,(a^2\cos^2\psi+b^2\sin^2\psi)+c^2\,\cos^2\vartheta\,,\nonumber\\
&&\omega_i\,\mathrm{d}x^i=V^{-1}\left((b^2-a^2)\,\sin\vartheta\,\sin\psi\,\cos\psi\,\mathrm{d}\vartheta+c^2\,\cos\vartheta\,\mathrm{d}\psi\right)\,,
\nonumber\\
&&\gamma_{\vartheta\vartheta}=a^2\,b^2\sin^2\vartheta+c^2\,\cos^2\vartheta\left(a^2\sin^2\psi+b^2\cos^2\psi\right)\,,\qquad\gamma_{tt}=\frac{V}{4}\,(abc)^2\,,
\\
&&\gamma_{\psi\psi}=c^2\sin^2\vartheta\,(a^2\cos^2\psi+b^2\sin^2\psi)\,,\qquad\gamma_{\vartheta\psi}=c^2(a^2-b^2)\sin\vartheta\,\cos\vartheta\,\sin\psi\,\cos\psi\,,\nonumber\\
&&\det g=V^{-2}\det\gamma\,,\qquad \det\gamma=\frac{\sin^2\vartheta}{4}\,V^2\,(abc)^4\,,\qquad \det g=\frac{\sin^2\vartheta}{4}\,(abc)^4\,,\nonumber
\end{eqnarray}
where $g$ and $\gamma$ are the metrics depicted in the line elements $\mathrm{d}\ell^2$ and $\mathrm{d}s^2$ respectively.
Ricci flatness of the four-dimensional metric introduces
the notion of the nut potential \cite{Gibbons:1979xm}
\begin{eqnarray}
\label{nut.eqn}
&&\delta(V^2\,\mathrm{d}\omega)=0\to \mathrm{d}(V^2\star_\gamma\mathrm{d}\omega)=0\to
V^2\star_\gamma\mathrm{d}\omega=\mathrm{d}b_{\rm nut}\Longrightarrow\nonumber\\
&&\partial_i\,b_{\rm nut}=\frac{1}{2}\sqrt{\gamma}\,V^2\varepsilon_i{}^{jk}\,\left(\partial_j\,\omega_k-\partial_k\,\omega_j\right)\,,
\quad \varepsilon_{t\,\vartheta\,\psi}=1\ ,
\end{eqnarray}
where the metric $\gamma_{ij}$ is used to define the superscripts and the covariant derivative.
Using the latter and   \eqref{DHL} we find the on-shell field
\begin{equation}
\label{nut.pot}
b_{\rm nut}=c\,(a+b-c)-\sin^2\vartheta\,\left((a-c)(a+c-b)-(a-b)(a+b-c)\,\sin^2\psi\right)\ .
\end{equation}
Combining the nut potential and $V$ we can built the fields
\begin{equation}
S_\pm=b_{\rm nut}\pm V\,.
\end{equation}
Using the latter we can define the quantity
\begin{equation}
\| \partial S_\pm\|^2=\gamma^{ij}\left(\partial_iS_\pm\right)\left(\partial_jS_\pm\right)\geqslant0\ ,
\end{equation}
where the strict inequality or equality is the criterion for a Killing vector to be rotational or translational,
respectively \cite{Boyer}. In our case
$\| \partial S_+\|^2=4$, so $\partial_\varphi$ is a rotational Killing vector\footnote{For the Belinsky--Gibbons--Page--Pope
metric \cite{Belinsky:1978ue};
which is a solution of the Lagrange system   \eqref{DHL},
$\xi_3$ turns out to be translational, i.e. $S_+=0$.}.
This enables us to define a coordinate $z$ as
\begin{equation}
\label{z-coord}
z=\frac{S_+}{\| \partial S_+\|}=
\frac{1}{2}\,\left(c(a+b)+\sin^2\vartheta\left(b(a-c)-c(a-b)\,\sin^2\psi\right)\right)\,,
\end{equation}
which will play the r\^ole of $z$ in the Toda frame \eqref{Boyer}.

The next step is to disentangle the $\mathrm{d}z^2$ from $\mathrm{d}s^2=\gamma_{ij}\,\mathrm{d}x^i\,\mathrm{d}x^j$
and this is done as follows
\be
\label{metric.split}
\mathrm{d}s^2-\mathrm{d}z^2=e_+\,e_-=4\mathrm{e}^{\Psi}\mathrm{d}q\,\mathrm{d}\bar q~,
\ee
with
\begin{equation}
\label{infgr3}
\begin{array}{rcl}
&& \displaystyle{e_+=e_+^t\,\mathrm{d}t+e_+^\vartheta\,\mathrm{d}\vartheta+e_+^\psi\,\mathrm{d}\psi~,\quad e_-=e_+^*~}\ ,
\crbig
&& \displaystyle{e_+^t=\frac{1}{2}abc\,\sin\vartheta\left(\cos\vartheta\left(a\cos^2\psi+b\sin^2\psi-c\right)+i\,(b-a)\,\sin\psi\,\cos\psi\right)}\, ,
\crbig
&& \displaystyle{e_+^\vartheta=ab\,\sin^2\vartheta+c\,\cos^2\vartheta\,\left(b\cos^2\psi+a\sin^2\psi\right)+i\,c\,(a-b)\,\cos\vartheta\,\sin\psi\,\cos\psi}\ ,
\crbig
&& \displaystyle{e_+^\psi=c\,\sin\vartheta\,\left((a-b)\cos\vartheta\sin\psi\,\cos\psi+i\left(a\cos^2\psi+b\sin^2\psi\right)\right)}\,,
\end{array}
\end{equation}where we made use of \eqref{DHL} and $2q=x+i\, y\,,2\bar q=2q^*=x-i\, y.$
Parametrising the Toda potential as $\mathrm{e}^\Psi=\mathrm{e}^{f+\bar f}$, we get the compatibility condition:\begin{eqnarray}
\label{compat.Toda}
&&e_+=2\mathrm{e}^f\,\mathrm{d}q\Longrightarrow\mathrm{d}\left(\mathrm{e}^{-f}\,e_+\right)=0\Longrightarrow \mathrm{e}^{2f}\,\mathrm{d}e_+=\frac{1}{2}\,\mathrm{d}\mathrm{e}^{2f}\wedge\,e_+\Longrightarrow\nonumber\\
&&\mathrm{e}^{2f}=2\left(a\,(b-c)\,\sin^2\vartheta+(b-a)\,c\,\left(\cos\vartheta\,\cos\psi-i\,\sin\psi\right)^2\right)~,\quad \mathrm{e}^{2\bar f}=\left(\mathrm{e}^{2f}\right)^*.
\end{eqnarray}
Using the latter, we find that
\begin{equation}
\label{Toda}
\mathrm{e}^{2\Psi}=4\left(\left(a\,(b-c)\,\sin^2\vartheta+c\,(a-b)\,\left(\sin^2\psi-\cos^2\vartheta\cos^2\psi\right)\right)^2+c^2\,(a-b)^2\,\cos^2\vartheta\,\sin^22\psi\right).
\end{equation}
We shall prove in Sec. \ref{VerifyToda}  that this is a solution of the continual Toda equation \eqref{Toda.Equation}. The latter equation is in agreement
with Olivier's result \cite{Olivier:1991pa} (Equation (40))
\begin{eqnarray}
\label{Toda.Olivier}
&&\mathrm{e}^{2f}=\sin^2\vartheta\,c\,(b-a)\,(\mu+\cosh\nu)\Longrightarrow \mathrm{e}^{\Psi}=\sin^2\vartheta\vert c\,(b-a)\vert\,\vert \mu+\cosh\nu\vert\,,\\
&&\mu=\frac{2ab-(a+b)c}{(b-a)c}=\frac{1+\kappa^2}{1-\kappa^2}\in\mathbb{R}\,,\qquad
\nu=2\left(\ln\tan\frac{\vartheta}{2}+i\,\psi\right)\in\mathbb{C}\,.
\end{eqnarray}
Straightforward calculation shows that the zeros of $\mathrm{e}^\Psi$ occur at:
\begin{equation}
 \label{AH.Toda.Zeros}
\psi=\pm\nicefrac{\pi}{2}\, \& \cos^2\vartheta=\kappa^2\in(0,1]\,,\qquad \vartheta=0,\pi\,,\qquad c=0\,,\quad a=b=c\, ,
\end{equation}
that is in four different cases.
We note that at these points there is a coordinate singularity of the metric $g$ in the Toda frame \eqref{Boyer}, since
\begin{equation}
\label{4ddet}
\det g=V^{-2}\det\gamma\Longrightarrow\det g=4V^{-2}\mathrm{e}^{2\Psi}=\left(\mathrm{e}^\Psi\,\partial_z\Psi\right)^2\,,
\end{equation}
which vanishes at the zeros of the Toda potential and its derivative.

Our next task is to compute $q,\bar q$ and complete the coordinate transformation. For this purpose we shall perform the
coordinate transformation
\begin{equation}
p=i\left(\psi+{\pi\ov 2}\right)+\ln\tan\frac{\vartheta}{2}\ .
\end{equation}
Using the latter, we find
\begin{equation}
\label{frame.new}\begin{array}{rcl}
&&\displaystyle{e_+=\ell_+^t\,\mathrm{d}t+\ell_+^\vartheta\,\mathrm{d}\vartheta+\ell_+^p\,\mathrm{d}p\,,\qquad e_-=e_+^*}\,,\crbig
&&\displaystyle{\ell_+^t=\frac{1}{2}\,abc\,\sin\vartheta\left(\cos\vartheta(b-c)+(a-b)\,\cosh p\,(\cosh p\cos\vartheta+\sinh p)\right)}\,,
\crbig
&&\displaystyle{\ell_+^\vartheta=\sin^2\vartheta\,\left(ab+c\,\left(a\,\sinh^2p-b\,\cosh^2p\right)\right)}\,,
\crbig
&&\displaystyle{\ell_+^p=c\,\sin\vartheta\,\left(a+(b-a)\cosh p\left(\cosh p+\cos\vartheta\sinh p\right)\right)}\,.
\end{array}
\end{equation}Indeed, one may verify that by using
\begin{equation}
\label{}
\begin{array}{rcl}
&&\displaystyle{ \sinh p = \sin\psi \cot\th + i {\cos\psi\ov \sin\th} \ ,\qquad \cosh p = -{\sin\psi\ov \sin\th} - i \cos\psi \cot\th}\ ,
\crbig
&&\displaystyle{ \cosh p \cos\th + \sinh p = i \cos\psi \sin\th \ ,\qquad\cosh p  + \cos\th\sinh p = -\sin\psi \sin\th}\,.
\end{array}
\end{equation}
The Toda potential reads
\begin{eqnarray}
&&\mathrm{e}^{2f}=\sin^2\vartheta\,P^2\,,\qquad \mathrm{e}^{2\bar f}=\left(\mathrm{e}^{2f}\right)^*\,, \\
&&P^2=2\,(ab+c(a\sinh^2p-b\cosh^2p))=2(a(b-c)+c(a-b)\cosh^2p)=c(b-a)\,(\mu-\cosh2p)\,,\nonumber
\end{eqnarray}
which satisfies $\mathrm{d}\left(\mathrm{e}^{-f}\,e_+\right)=0$ as it should.
The next step is to consider an ansatz for the coordinate $q$ \cite{Olivier:1991pa,Finley:2010hs}
\begin{equation}
\label{qans}
2q=-\frac{1}{2}\,P\cos\vartheta+W(t,p)\ .
\end{equation}
It is easy to check, that for this choice we retrieve the $\ell_+^\vartheta$ component of  \eqref{frame.new}.
This enables us to write a system of first order partial differential equations for $W(t,p)$, namely:
\begin{equation}
\label{pdeW}
P\,\frac{\partial W}{\partial t}=\frac{1}{2}\,abc(a-b)\sinh p\cosh p\,, \qquad
P\,\frac{\partial W}{\partial p}=c\,(b\cosh^2p-a\sinh^2p)\,,
\end{equation}
which can be easily proved through  \eqref{DHL} to be compatible, i.e.
$\left[\frac{\partial }{\partial t},\frac{\partial }{\partial p}\right]W(t,p)=0.$
Integrating the second one, we find
\begin{eqnarray}
\label{W.partial.int.p}
&&\sqrt{2}\,W(t,p)=\frac{ab}{\sqrt{a(b-c)}}\,\int\frac{\mathrm{d}p}{\sqrt{1-\kappa'^2\cosh^2p}}-\sqrt{a(b-c)}\,\int\mathrm{d}p\,\sqrt{1-\kappa'^2\cosh^2p}+\xi(t)
\nonumber\\
&&\Longrightarrow\sqrt{2}\,W(t,p)=i\left\{\sqrt{b(a-c)}\,E\left(i\,p,i\frac{\kappa'}{\kappa}\right)-
\frac{ab}{\sqrt{b(a-c)}}\,F\left(i\,p,i\frac{\kappa'}{\kappa}\right)\right\}+\xi(t)\,,\\
&&\kappa^2=\frac{b(c-a)}{a(c-b)}\,,\qquad \kappa'^2=1-\kappa^2=\frac{c(a-b)}{a(c-b)}\,,\nonumber
\end{eqnarray}
where $\xi(t)$
is an arbitrary function of time and $F(t,k),E(t,k)$ are incomplete elliptic integrals of the first and second type
\begin{equation}
\label{elliptic.integrals}
F(t,k)=\int_0^t\mathrm{d}y\frac{1}{\sqrt{1-k^2\sin^2y}}\,,\qquad E(t,k)=\int_0^t\mathrm{d}y\sqrt{1-k^2\sin^2y}\,.
\end{equation}
To specify this function, we shall compute the partial derivative of $W(t,p)$ with respect to $t$, use \eqref{DHL} and compare it with the first equation of \eqref{pdeW}:
\begin{equation}
\label{Wt}
\frac{\partial W}{\partial t}=\frac{1}{2P}\,abc(a-b)\sinh p\cosh p+\frac{\mathrm{d}\xi(t)}{\mathrm{d}t}\Longrightarrow\frac{\mathrm{d}\xi(t)}{\mathrm{d}t}=0\Longrightarrow\xi(t)={\rm constant}\ ,
\end{equation}
which additional constant we can always set to zero.
Combining   \eqref{z-coord}, \eqref{Toda}, \eqref{qans}, \eqref{W.partial.int.p} and \eqref{Wt} we retrieve the full coordinate transformation
from $(t,\vartheta,\psi)$ to $(q,\bar q,z)$ which was given in   \eqref{coordzq}.

Finally, we mention that the explicit solution of \eqref{DH} was not used for the integration of $q=q(t,\vartheta,p)$
in contrast with the derivation which was performed in App. C of \cite{Finley:2010hs}.

\subsection{Verification of the solution}
\label{VerifyToda}
So, we have the desired coordinate transformation $y^\alpha=(q,\bar q,z)$ and the Toda potential $\Psi$ as functions of $x^i=(t,\vartheta,\psi).$
The scope of this section is to prove that $\Psi$ satisfies the continuoual Toda equation \eqref{Toda.Equation}. Taking into account that
the expressions are transcendental, we shall use the chain rule
\begin{equation}
\frac{\partial\,\Psi}{\partial y^\alpha}=\frac{\partial\,x^i}{\partial y^\alpha}\,\frac{\partial\,\Psi}{\partial x^i}\,.
\end{equation}
However, the expressions we have, involve the new coordinates $y^\alpha$ as functions of the old ones $x^i$ and so, we have to compute the Jacobian matrix
of the coordinate transformation\footnote{Note that the $\det\gamma$  transforms as it should: i.e. a scalar density of weight $(h_q,h_{\bar q})=(1,1)$; \\
$\displaystyle\det\gamma=JJ^*\det\widetilde\gamma=\frac{\sin^2\vartheta}{4}\,V^2\,(abc)^4$.}
\begin{equation}
\label{Jacobian}
x^i\mapsto y^\alpha\,,\qquad J^\alpha{}_i=\frac{\partial y^a}{\partial x^i}\,,\qquad
\det J=\frac{i}{4}\,(abc)^2\,\sin\vartheta\,V\,\mathrm{e}^{-\Psi}\,,
\end{equation}
whose expression can be found with the use of \eqref{z-coord},\eqref{metric.split} and \eqref{DHL}.
Inverting this matrix we can compute the derivatives with respect to $y^\alpha$ as follows
\begin{equation}
\label{chain.rule}
\frac{\partial}{\partial y^\alpha}=(J^{-1})^i{}_\alpha\,\frac{\partial}{\partial x^i}\,.
\end{equation}
Using the latter and Eqs. \eqref{Toda},\eqref{DHL} we find a simple expression for the partial derivatives of $\Psi$ with respect to $z$
\begin{equation}
\label{partialz.Toda}
\frac{\partial\Psi}{\partial z}=\frac{1}{2\mathrm{e}^{2\Psi}}\frac{\partial \mathrm{e}^{2\Psi}}{\partial z}=\frac{2}{V}\,,\qquad
\frac{\partial \mathrm{e}^{2\Psi}}{\partial z}=4V^{-1}\mathrm{e}^{2\,\Psi}\,,\qquad \frac{\partial^2\mathrm{e}^{\Psi}}{\partial z^2}=
2\,\mathrm{e}^{\Psi}\,V^{-2}\left(2-\frac{\partial\, V}{\partial z}\right)\,,
\end{equation}
which is in agreement with  \eqref{Boyer}.
To compute the partial derivatives with respect to $q,\bar q$ of $\Psi$, it is convenient  to consider the product and the ratio of
$\mathrm{e}^{2f}$ and $\mathrm{e}^{2\bar f}$ as in \cite{Finley:2010hs}
\begin{equation}
\Pi=\mathrm{e}^{2\Psi}=\mathrm{e}^{2f+2\bar f}\,,\qquad R=\mathrm{e}^{2f-2\bar f}\,.
\end{equation}
To proceed, it is helpful to rewrite the partial derivatives with respect to $q$ and $\bar q$ as
\begin{equation}
\frac{\partial}{\partial q}=\mathrm{e}^f\, {\cal Q}\,,\qquad \frac{\partial}{\partial \bar q}=\mathrm{e}^{\bar f}\,\bar {\cal Q}\,,
\end{equation}
where ${\cal Q},\bar {\cal Q}$ should be considered as first order differential operators acting on the right and obeying Leibniz rule.
Using these operators, we can write the second partial derivative of $\Psi$ with respect to
$q$ and $\bar q$ as follows
\begin{eqnarray}
&&\frac{\partial^2\Psi}{\partial q\partial\bar q}=\frac{1}{2}\,\mathrm{e}^\Psi\left(\frac{{\cal Q}\bar {\cal Q}\,\Pi}{\Pi}-\frac{3}{4}\,\frac{{\cal Q}\Pi\,\bar {\cal Q}\Pi}{\Pi^2}-\frac{1}{4}\frac{{\cal Q}R\,\bar {\cal Q}\Pi}{\Pi\,R}\right)\,,\\
&&\frac{\partial^2\Psi}{\partial \bar q\partial q}=\frac{1}{2}\,\mathrm{e}^\Psi\left(\frac{\bar {\cal Q}\, {\cal Q}\Pi}{\Pi}-\frac{3}{4}\,\frac{{\cal Q}\Pi\,\bar {\cal Q}\Pi}{\Pi^2}+\frac{1}{4}\frac{{\cal Q}\Pi\,\bar {\cal Q}R}{\Pi\,R}\right)\,.
\end{eqnarray}
Commuting of the partial derivatives, provides the following constraint\footnote{
We use the standard notations for commutators and anticommutators
$$[A,B]=AB-BA\ ,\qquad \{A,B\}=AB+BA\ .$$ }
\begin{equation}
\label{2derconst}
\frac{4}{\Pi}[{\cal Q},\bar {\cal Q}]\Pi=\frac{{\cal Q}\Pi\,\bar {\cal Q}R+\bar {\cal Q} \Pi\,{\cal Q}R}{\Pi\,R}\,,
\end{equation}
which is satisfied after a lengthy but straightforward computation. Thus, we can write the second partial derivative of $\Psi$ with respect to
$q$ and $\bar q$ in a symmetric way
\begin{equation}
\label{2derPsi}
\frac{\partial^2\Psi}{\partial q\partial\bar q}
=\frac{1}{4}\,\mathrm{e}^\Psi\left(\frac{\{{\cal Q},\bar {\cal Q}\}\Pi}{\Pi}-\frac{3}{2}\,\frac{{\cal Q}\Pi\,
\bar {\cal Q}\Pi}{\Pi^2}+\frac{1}{4}\frac{{\cal Q}\Pi\,\bar {\cal Q}R-{\cal Q}R\,\bar {\cal Q}\Pi}{\Pi\,R}\right)\ .
\end{equation}
Finally, plugging \eqref{partialz.Toda} and \eqref{2derPsi} in the continual Toda equation \eqref{Toda.Equation}, we find after a
lengthy but straightforward computation,
that it is indeed satisfied.

\section{Modular forms and elliptic integrals}

\label{modfor}

We collect here some conventions for the modular forms and theta functions
used in the main text. General results and properties of these objects can be found in~\cite{serre,Koblitz}.

Introducing  $q=\mathrm{e}^{2i\pi \tau}, \ \tau\in \mathbb{C}$, we first define
\be
\label{A1}
\eta(\tau) =q^{\nicefrac{1}{24}} \prod_{n=1}^\infty\left(1-q^n\right)\ ,
\qq E_2(\tau)=\frac{12}{i\pi}\frac{\mathrm{d}\ln \eta }{\mathrm{d}\tau}  \ ,
\ee
as the Dedekind function and the weight-two quasimodular form.
The Dedekind function has the modular transformations
 \begin{equation}
\eta(\tau+1)=\mathrm{e}^{i\pi/12}\,\eta(\tau)\,,\qquad\eta(-1/\tau)=\sqrt{-i\tau}\, \eta(\tau)\ .
 \end{equation}
In addition note the Jacobi theta functions.
 \begin{equation}
  \theta_2(\tau)=\sum_{p\in \mathbb{Z}} q^{\nicefrac{1}{2}\left(
  p+\nicefrac{1}{2}\right)^2},\quad
  \theta_3(\tau)=\sum_{p\in \mathbb{Z}} q^{\nicefrac{p^2}{2}},\quad
  \theta_4(\tau)=\sum_{p\in \mathbb{Z}} (-1)^p\, q^{\nicefrac{p^2}{2}}\ .
\end{equation}
These have many remarkable properties. We quote here the Jacobi identity
\be
\label{theta4}
\theta_2^4-\theta_3^4+\theta_4^4=0\
\ee
and
\be
E_2(i)=\frac{3}{\pi}\ .
\label{E2i}
\ee
Their transformation properties under a M\"obius transformation are
 \begin{equation}
\begin{cases}
  \theta_2(\tau+1)= \sqrt{i}\, \theta_2(\tau)\\
  \theta_3(\tau+1)= \theta_4(\tau)
 \\
  \theta_4(\tau+1)=  \theta_3(\tau)
\end{cases}{\rm and}\quad
\begin{cases}
  \theta_2(-\nicefrac{1}{\tau})= \sqrt{-i\tau}\, \theta_4(\tau)\\
  \theta_3(-\nicefrac{1}{\tau})= \sqrt{-i\tau}\, \theta_3(\tau)
 \\
  \theta_4(-\nicefrac{1}{\tau})= \sqrt{-i\tau}\, \theta_2(\tau)
\end{cases}
\label{modtheta}
\end{equation}
and
 \begin{equation}
E_2(\tau+1)=E_2(\tau)\ ,\qq
E_2(-\nicefrac{1}{\tau})=\frac{6\tau}{i\pi}+\tau^2 E_2(\tau)\ .
\end{equation}
The last relation can be used to provide a proof of \eqn{E2i}.

The Jacobi functions and the weight-two quasimodular form are related to the complete elliptic integrals
of the first and second kind, defined respectively as special values of the incomplete integrals \eqref{elliptic.integrals}
 \begin{equation}
 \label{cel1}
 K(\kappa)=F\left(\frac{\pi}{2},\kappa\right) \ , \qquad
 E(\kappa)=E\left(\frac{\pi}{2},\kappa\right)\ ,
 \end{equation}
 which satisfy the Legendre relation
 \begin{equation}
\label{Legendre}
 K(\kappa)E(\kappa')+E(\kappa)K(\kappa')-K(\kappa)K(\kappa')=\frac{\pi}{2}\ .
 \end{equation}
Setting  $\tau$ as the elliptic modulus
 \begin{equation}
  \label{elmod}
 \tau=i\frac{K(\kappa')}{K(\kappa)}\ ,
\end{equation}
we obtain
\begin{equation}
 \label{elnorm}
\kappa=\frac{\theta_2^2(z)}{\theta_3^2(z)}\ , \qq \kappa'=\frac{\theta_4^2(z)}{\theta_3^2(z)}
\quad \Longrightarrow\quad \kappa^2 + \kappa'^2=1
\end{equation}
and
\begin{eqnarray}
 \label{model}
K(\kappa)&=&\frac{\pi}{2}\theta_3^2(\tau)\ ,
\\
\label{id1}
E(\kappa)&=&K(\kappa) +\frac{\pi^2}{12K(\kappa)}\left(E_2(\tau)-\theta_2^4(\tau)-\theta_3^4(\tau)\right)\ ,
\end{eqnarray}
or equivalently
\begin{equation}
 \label{id2}
E(\kappa)K(\kappa)-\kappa'^2K^2(\kappa) =\frac{\pi^2}{12}\left(E_2(\tau)+\theta_2^4(\tau)-\theta_4^4(\tau)\right)\ .
\end{equation}
Notice finally the useful inequality
\begin{equation}
 \label{Cineq}
0<E(\kappa')-\kappa^2K(\kappa')<\frac{\pi}{2K(\kappa)}\ .
\end{equation}

\boldmath
\section{Extra Killing vector and electrostatics}
\unboldmath
\label{electro.Toda}

We will consider solutions which have an additional $U(1)$ isometry.
Not much is really new material but we believe that the presentation is novel.
It can be read independently from the rest of the paper.

In the case of backgrounds with an extra Killing vector field $\del_\b$ and an associated $U(1)_\beta$ symmetry (compact or non-compact)
we can recast the solutions of the continual Toda equation into a problem of electrostatics with appropriate boundary conditions.
From the point of view of the auxiliary four-dimensional problem \eqn{GH} with $\del_\varphi$ a rotational Killing vector, the new Killing vector
$\del_\b$ could be either translational or rotational. It could correspond to one of the Cartesian coordinates $x$ and $y$,
or to the angle $\b$ when we use polar coordinates $(r,\b)$ instead of the Cartesian ones $(x,y)$. The mapping to the electrostatic problem
was done in \cite{Ward:1990qt} when the extra Killing isometry was one the Cartesian coordinates.
Both cases can be treated in parallel as they are mapped on one each other.
Indeed, consider the continual Toda Eq. \eqn{Toda2} for solutions which are independent of the variable $y$.
Then consider the transformation
\be
\Psi \mapsto \Psi + 2\ln r\ ,\qq x\mapsto \ha \ln r\ ,\qq z \mapsto {z\ov 2}\ ,
\ee
Then, thanks to $r\partial_r\left(r\partial_r\ln r\right)= r\,\delta(r)=0$ and using \eqref{Toda2},
the new function $\Psi(r,z)$ satisfies
\be
\label{cyl}
\frac{1}{r}\partial_r\left(r\partial_r\Psi\right)+\partial^2_z \mathrm{e}^\Psi=0\ ,
\ee
which is the continual Toda equation \eqn{Toda2} in cylindrical polar coordinates with azimuthal symmetry.

Here we present some necessary formulae for the case corresponding to an isometry along the polar angle in the $x$-$y$ plane, Eq. \eqref{cyl}.
We let
\be
\ln r = \partial_\eta \Phi \equiv \Phi'\ ,\qquad z = \rho\partial_\rho \Phi \equiv \dot \Phi \ ,\qquad \rho=r \mathrm{e}^{\Psi(r,z)/2}\ ,
\label{er0}
\ee
where $\Phi=\Phi(\rho,\eta)$ satisfies the scalar Laplacian in the cylindrical coordinates $(\rho,\eta)$
\begin{equation}
\label{Laplacian}
\frac{1}{\rho}\partial_\rho\left(\rho\partial_\rho\Phi\right)+\partial^2_\eta\Phi=0\quad \Longrightarrow  \quad
\ddot \Phi + \r^2 \Phi^{\prime\prime} = 0\ .
\end{equation}
Given the above change of variable for $\r=\r(r,z)$ one may compute the other variable $\eta=\eta(r,z)$ from the exact differential
\be
\mathrm{d}\eta = - {\r\ov r} \del_z \r\ \mathrm{d}r + {r\ov \r} \del_r \r\ \mathrm{d}z\ .
\label{diffH}
\ee
In addition the potential $\Phi$  is computed form the following exact differential
\be
\mathrm{d}\Phi = \left(-\frac{\rho}{r}\,\partial_z\rho\,\ln r+\frac{z}{\rho}\partial_r\rho\right)\mathrm{d}r +
\left(\frac{z}{\rho}\,\partial_z\rho+\frac{r}{\rho}\,\partial_r\rho\,\ln r\right)\mathrm{d}z\  .
\label{diffF}
\ee

The boundary is at $z=0$, which implies the boundary condition
\be
\del_\r \Phi \big |_{\eta=0} = 0\ .
\ee
and a charge density $\l(\eta)$ along the $\eta$-axis. We may take $\Phi(\r,0)=0$ and
the boundary condition can be satisfied by considering an image line with opposite charge below the plane, i.e. $\l(-\eta)=-\l(\eta)$.
The above define an electrostatic problem.
Due to the line density the right hand side of the Laplace equation \eqn{Laplacian} should be replaced by
$\displaystyle \l(\eta) {\d(\r)\ov \r}$.
From Gauss' law applied on a cylinder with small radius and small height which does not cross the $\eta=0$ plane
we have that for the potential and the radial component of the electric field
\be
\Phi \simeq  \l \ln \r +\cdots \quad \Longrightarrow \quad  E_\r = - \del_\r \Phi \simeq  - {\l \ov  \r} +\cdots\ .
\label{begrs0}
\ee
If we may easily compute the behaviour of the potential $\Phi$ near $\r=0$, then this is a practical way to read off the charge line density $\l(\eta)$.
Another way is to write the second of the above as
\be
\l(\eta) \equiv \r\del_\r \Phi\big |_{\r =0} = z(\r=0,\eta) \ .
\ee
Therefore if we know the explicit form of the coordinate change from $(r,z)\mapsto (\r,\eta)$, it might be more practical to use the above
formulae.
Returning to the computation of the potential $\Psi$ and assuming that the space is free of boundaries we have that
\be
\Phi(\r,\eta)   = -\ha \int_0^\infty \mathrm{d}\eta'  \l(\eta') G(\r,\eta;\eta')\ ,
\label{gfikp}
\ee
where the Green's function is
\be
G(\rho,\eta;\eta')=\frac{1}{\sqrt{\rho^2+(\eta-\eta')^2}}-\frac{1}{\sqrt{\rho^2+(\eta+\eta')^2}}
\ee
and satisfies the standard Green's equation with Dirichlet boundary condition
\be
\nabla^2 G(\r,\eta;\eta') = {1\ov \r} \del_\r (\r \del_\r G) + \del_\eta^2 G =
 -{2\ov \r} \d(\r) \d(\eta-\eta')\ ,\qq G(\r,\eta,\eta')\big |_{\eta=0}\ .
\ee

The electrostatics picture of the Toda equation implies a smearing procedure has occurred so that an extra
$U(1)$ isometry develops.
That introduces the limitation that the length-scales for which a solution is trustworthy should be larger than the smearing length, i.e. $\rho_{\rm sm}$.
In addition the charge line density should obey some consistency conditions so to avoid coordinate singularities \cite{Gaiotto:2009gz}.
Finally, these profiles only tell us about the behaviour of the solution around $\rho=0$. At larger
distances, say for $\rho>\rho_{U(1)}$, it should be replaced by a solution of the continual Toda without a $U(1)_\beta$ isometry.
Thus the electrostatics picture is valid when $\rho_{\rm sm}\ll \rho\ll \rho_{U(1)}$ and for specific forms of the line density.
An exception is the Maldacena--N\'u\~nez solution \cite{Maldacena:2000mw} which  is an unsmeared solution.

\subsection{Maldacena--N\'u\~nez solution}
\label{sec.electro.MN}
The first paradigm we are going to tackle is the Toda potential of the Maldacena--N\'u\~nez solution \cite{Maldacena:2000mw}
\begin{equation}
\label{Toda.MN}
\mathrm{e}^\Psi=4\frac{N^2-z^2}{\left(1-r^2\right)^2}\,,\qquad z\in[0,N]\, ,\qquad r\in[0,1]\, ,
\end{equation}
which is the Toda potential of the singular Eguchi--Hanson metric for the rotating Killing vector.\footnote{The corresponding Gibbons--Hawking four-dimensional geometry has an
 anti-self-dual Riemann two-form. It is a Ricci flat space
whose Kretschmann scalar reads $R_{\kappa\lambda\mu\nu}\,R^{\kappa\lambda\mu\nu}=\frac{24N^4}{z^6}$, so that
the geometry is singular at $z=0$.
The Toda potential for the regular Eguchi--Hanson metric has been computed in \cite{Bakas:1996gf}. The expression \eqn{Toda.MN} is
an analytic continuation of equation (3.10) in \cite{Bakas:1996gf}.}
Form \eqn{er0} and \eqref{Toda.MN} we find that
\begin{equation}
\label{rho.MN}
\rho= \frac{2r\,\sqrt{N^2-z^2}}{1-r^2}\geqslant0\, .
\end{equation}
Using \eqn{diffH} we compute
\be
\label{electro.MN}
\eta=\frac{1+r^2}{1-r^2}\,z\geqslant0\ .
\ee
One can invert the above expressions but the resulting expressions for $r(\r,\eta)$ and $z(\r,\eta)$ are not very illuminating.
The electrostatic potential can be computed using \eqn{diffF}. One finds that
\be
\Phi =z-N \tanh ^{-1}\frac{z}{N}+\frac{1+r^2}{1-r^2}\,z\ln r\ .
\ee
Moreover, its charge density at $\rho=0$ which corresponds to $r=0$ or  $z=N$ reads
\begin{equation}
\label{density.MN}
\lambda(\eta)= \left\{
 \begin{array}{lr}
       \eta\,, &\quad 0\leqslant\eta\leqslant N\,,\\
       N\,, &\quad \eta\geqslant N\ .
     \end{array}
   \right.
\end{equation}
This line density can be also computed by expanding $\Phi$ near $r=0$ and near $z=N$, corresponding to expanding near $\r=0$.
In addition, for consistency one may verify that this electrostatic potential can be directly evaluated by using \eqn{gfikp}.

\boldmath
\subsection{$\text{AdS}_7\times S^4$ solution}
\label{electro.AdS}
\unboldmath
The next paradigm concerns the space $\text{AdS}_7 \times S^4$ which can be cast \cite{Gaiotto:2009gz} in the form of \eqn{11dsolution}
with
\begin{equation}
\label{Toda.AdS}
\mathrm{e}^{\Psi}=\coth^2\zeta\ ,
\end{equation}
where the variables $(r,z)$ are expressed in terms of the auxiliary variables $\zeta\geqslant 0$ and $\vartheta \in [0,\pi/2]$ as
\be
r=\sinh^2\zeta\sin\vartheta\ ,\qquad z=\cosh^2\zeta\cos\vartheta\ .
\label{impliir}
\ee
Then using also \eqn{er0} we immediately see that
\be
\rho=\frac{1}{2}\,\sinh2\zeta\sin\vartheta\ .
\ee
Then from \eqn{diffH}, \eqn{diffF} and \eqn{impliir} we find that
\be
\eta=\frac{1}{2}\cosh2\zeta\cos\vartheta\
\ee
and that
\be
\Phi=\frac{1}{2}\left(\cos\vartheta\left(1+\cosh2\zeta\ln r\right)+\ln\tan\frac{\vartheta}{2}\right)\ .
\ee
Finally, its charge density at $\rho=0$, corresponds either to $\zeta=0$ or $\vartheta=0$, reads
\begin{equation}
\label{density.AdS}
   \lambda(\eta) = \left\{
     \begin{array}{cc}
  & 2\eta\, ,\qq 0 \leqslant \eta\leqslant \ha \,,\\
   & \eta+\frac{1}{2}\, ,\qq \eta\geqslant\frac{1}{2}\,.
     \end{array}
   \right.
\end{equation}
Again, this line density can be also computed by expanding $\Phi$ near $\zeta=0$ and near $\th=0$, corresponding to an expansion near $\r=0$.
In addition, this electrostatic potential can be directly evaluated by using \eqn{gfikp} with the above line density.

\subsection{The instanton picture}
It is interesting to further explore the nature of the four-dimensional instantonic solution
corresponding to the above Toda potential.
To identify the type of instanton, it is useful to write the corresponding Gibbons--Hawking metric
for the Toda frame of \eqref{Toda.AdS}
\begin{equation}
\label{MG.Instanton.varphi}
\begin{array}{rcl}
 &&\displaystyle{\mathrm{d}\ell^2=V(\mathrm{d}\varphi+\omega)^2+V^{-1}\left(\mathrm{d}z^2+\mathrm{e}^\Psi\left(\mathrm{d}r^2+r^2\mathrm{d}\beta^2\right)\right)}\,,\crbig
 &&\displaystyle{\omega =-\frac{1}{2}r\,\partial_r\Psi\,\mathrm{d}\beta\,,\qquad V=\frac{2}{\partial_z\Psi}}\,.
\end{array}
\end{equation}
It turns out that this metric has two rotational Killing vectors $(\partial_\varphi,\partial_\beta)$ and  a linear combination of them given by
$\partial_{\tilde\beta}=\partial_\beta-\partial_\varphi$ is translational since \eqref{selfKilling} is satisfied (with the minus sign)
in agreement with a theorem in \cite{Gibbons:1987sp}.
Explicitly we have that
\begin{equation}
 \label{MG.Instanton.beta.phi}
 \begin{array}{rcl}
 &&\displaystyle{\mathrm{d}\ell^2= V_f(\mathrm{d}\tilde\beta+\omega_f)^2+V_f^{-1}
 \left(V^{-1} V_f(\mathrm{d}z^2+\mathrm{e}^\Psi \mathrm{d}r^2)+r^2\mathrm{e}^\Psi\mathrm{d}\varphi^2\right)}\,,\crbig
 &&\displaystyle{V_f=V(\omega_\beta-1)^2+r^2\mathrm{e}^\Psi\,V^{-1}\,,\qquad
 \omega_f=\frac{(\omega_\beta-1) V}{V_f}\mathrm{d}\varphi}\,.
\end{array}
\end{equation} The three-dimensional flat metric $\gamma_{ij}$ and the potential $V_f^{-1}$ obey
\begin{equation}
\label{}
\begin{array}{rcl}
&&\displaystyle{\gamma_{\zeta\zeta}=4\gamma_{\vartheta\vartheta}=\frac12(\cosh4\zeta-\cos2\vartheta)\,,\qquad \gamma_{\varphi\varphi}=\frac14\sin^2\vartheta\sinh^22\zeta}\,,\crbig
&&\displaystyle{V_f^{-1}=\frac{4\cos\vartheta}{\cos2\vartheta-\cosh4\zeta}\,,\qquad
R_{ij;kl}=0\,,\qquad \nabla^2_\gamma V_f^{-1}=0}\,.
\end{array}
\end{equation}
It is easy to check that $\gamma_{ij}$ is the induced metric in prolate spheroidal coordinates,
where the coordinate transformation reads
 \begin{equation}
 x=\frac12\sinh2\zeta\sin\vartheta\cos\varphi\,,\quad y=\frac12\sinh2\zeta\sin\vartheta\sin\varphi\,,\quad
z=\frac12\cosh2\zeta\cos\vartheta\,,
\end{equation}
with $\zeta\geqslant0\,,\ \vartheta\in[0,\pi]\,,\ \varphi\in[0,2\pi]$.
Using this parametrisation we find that the potential $V_f^{-1}$ can be written as
\begin{equation}
V_f^{-1}=\frac{1}{2}\left(\frac{1}{\vert\vec r+\vec r_0\vert}-\frac{1}{\vert\vec r-\vec r_0\vert}\right)\,,
\end{equation}
where $\vec r(\zeta,\vartheta,\varphi)=(x,y,z)$ and $\vec r_0=\vec r(0,0,\varphi)=(0,0,1/2)$.
Comparing the latter with the generic form of localised solutions of the Laplace equation \eqref{Trans.Laplace},
we find a two centre Eguchi--Hanson with moduli parameters $\pm 1/2$ and $\varepsilon=0$ \cite{Gibbons:1979zt}.
This is however a singular solution in agreement with our general result that solutions
of the continual Toda equation which are appropriate for constructing eleven-dimensional backgrounds lead to singular
four-dimensional instantonic solutions.

\end{document}